\documentclass[manuscript,nonacm]{acmart}
\AtBeginDocument{%
  }

\acmDOI{}

\acmJournal{TOSEM}

\usepackage{longtable}
\usepackage{array}
\usepackage{ltcaption}
\usepackage{booktabs}
\usepackage{rotating}
\usepackage{tabularray}
\usepackage{ragged2e}
\usepackage{orcidlink}

\usepackage{tcolorbox}
\newtcolorbox{textbox}[1]{
    sharp corners,
    boxsep = 0mm,
    toptitle = 2mm,
    lefttitle = 0mm,
    colframe=gray!5,
    colback=gray!5,
    title={\rule[-2pt]{4.5pt}{10pt}\hspace*{1.5mm}#1},
    coltitle=black,
    halign = flush left,
    parbox = false,
}

\usepackage[table]{xcolor}
\definecolor{reqcolor}{RGB}{230,245,255}      
\definecolor{ctxcolor}{RGB}{245,235,255}      
\definecolor{syscolor}{RGB}{235,255,235}      

\definecolor{omittederrcolor}{RGB}{255,248,220}   
\definecolor{falseprederrcolor}{RGB}{255,230,230} 
\definecolor{missederrcolor}{RGB}{232,245,250}    

\usepackage{listings}
\lstdefinestyle{pythonstyle}{
    language=Python,
    basicstyle=\ttfamily\small,      
    keywordstyle=\color{blue},
    stringstyle=\color{teal},
    commentstyle=\color{gray},
    showstringspaces=false,
    breaklines=true,
    frame=single
}

\begin{document}

\title{ReqSeek: Transformer-Based Automatic Requirements Identification for Building Large Requirements Datasets}

\author{Mohammad Kasra Habib\orcidlink{0000-0002-1272-9873}}
\email{kasra.habib@tum.de}
\orcid{0000-0002-1272-9873}
\affiliation{%
  \institution{Technical University of Munich}
  \city{Heilbronn}
  \country{Germany}
}

\author{Umm-e-Habiba\orcidlink{0000-0001-8953-9624}}
\email{umme.habiba@tum.de}
\orcid{0000-0001-8953-9624}
\affiliation{
  \institution{Technical University of Munich}
  \city{Heilbronn}
  \country{Germany}
}

\author{Daniel Graziotin\orcidlink{0000-0002-9107-7681}}
\email{graziotin@uni-hohenheim.de}
\orcid{0000-0002-9107-7681}
\affiliation{
  \institution{University of Hohenheim}
  \city{Stuttgart}
  \country{Germany}}

\author{Stefan Wagner\orcidlink{0000-0002-5256-8429}}
\email{stefan.wagner@tum.de}
\orcid{0000-0002-5256-8429}
\affiliation{
  \institution{Technical University of Munich}
  \city{Heilbronn}
  \country{Germany}
}

\renewcommand{\shortauthors}{Habib et al.}

\begin{abstract}
Large, diverse, and high-quality datasets are significant for advancing AI in requirements engineering, yet automated identification of requirements from Software Requirement Specifications (SRS) remains a challenge. Automatic Requirements Identification (ARI) methods demonstrate limited generalization to unseen documents across heterogeneous SRSs.

This work presents ReqSeek, a transformer-based, domain-agnostic ARI model for recognizing ISO 29148-style requirements in SRS documents, supporting dataset construction. We curated the Automatic Requirements Identification Dataset (ARID), comprising 2,396 training instances from 23 heterogeneous SRSs, and fine-tuned four transformer architectures. RoBERTa was selected as the base model for ReqSeek and achieved 95\% precision, recall, $F_1$, and $F_{.89}$ on the held-out evaluation set. Exploratory human–model benchmarking showed performance comparable to humans while requiring substantially less prediction time than manual annotation. Post-hoc evaluation on out-of-domain (OoD) SRSs provided evidence of requirement identification beyond ARID. Robustness analysis demonstrated ReqSeek's insensitivity to minor linguistic variations while maintaining sensitivity to major ISO 29148-style violations. We applied ReqSeek to construct the Software Requirements Dataset (SwaRD), identifying 10,201 ISO 29148-style requirements from 231 industrial SRSs. SwaRD aligns with textual and semantic properties of human-curated datasets while exceeding them in scale and domain diversity. Our results suggest that transformer-based ARI can provide precise and generalizable requirements identification.

\end{abstract}

\ccsdesc[500]{Computing methodologies~Supervised learning by classification}
\ccsdesc[500]{Computing methodologies~Natural language processing}
\ccsdesc[500]{Software and its engineering~Requirements analysis}
\ccsdesc[500]{Information systems~Language models}

\keywords{Software Engineering, Requirements Engineering, Automatic Requirements Identification, Requirements Classification, Transformer-Based Models, Requirements Dataset, ReqSeek, ARID, SwaRD}

\received{21 January 2026}
\received[revised]{12 April 2026}
\received[accepted]{29 October 2026}

\maketitle

\section{Introduction}
The \textbf{lack of large-scale datasets} for advancing requirements engineering research is a commonly highlighted problem \cite{ferrariNaturalLanguageRequirements2017, alhoshanZeroShotLearningApproach2022, liMachineLearningRequirements2022a, zarourRequirementsEngineeringPractices2025, dengNFRNetLTImprovingAccuracy2023}. This limitation is not due to the unavailability of Software Requirements Specifications (SRS), but rather to the \textbf{difficulty of extracting requirements} from them \cite{liMachineLearningRequirements2022a, izharBridgingPrecisionComplexity2025}. This challenge stems from the complexity of SRS documents \cite{falknerIdentifyingRequirementsRequests2019}, which are often non-machine-readable and intermix requirements with supplementary descriptions. As a result, requirement identification is practical only for a limited number of documents when performed manually. Consequently, assembling a requirements dataset from multiple SRS documents with diverse formats is highly time-consuming and labor-intensive, particularly when requirements content must be precisely separated from non-requirements to preserve dataset quality. Addressing this gap is therefore imperative.

\textbf{\textit{Our goal is to develop a domain-agnostic model for Automatic Requirements Identification (ARI) that identifies requirements formulated in compliance with ISO 29148 and supports the systematic construction of large-scale requirements datasets for AI applications in requirements engineering research.}}

This emphasis on ISO 29148-style requirements builds on existing ARI research \cite{falknerIdentifyingRequirementsRequests2019, abualhaijaMachineLearningBasedApproach2019, ivanovExtractingSoftwareRequirements2022a}. ISO 29148 provides syntactic and semantic guidance for writing explicit, verifiable system obligations or provisions, making ISO 29148-style requirements practically important for requirements specification, verification, and traceability \cite{ISOIECIEEE}. Prior ARI studies have commonly operationalized the identification of ISO 29148-style requirements as binary classification, distinguishing binding \textit{(1) requirements} from \textit{(2) auxiliary content} such as explanations, examples, or references \cite{winklerAutomaticClassificationRequirements2016a}. These studies applied ARI to support downstream activities such as minimizing oversight risk \cite{falknerIdentifyingRequirementsRequests2019}, facilitating risk assessments \cite{bashirRequirementNotThat2023, winklerAutomaticClassificationRequirements2016a}, and enabling traceability and validation for software reuse \cite{hayesWekaMeetsTraceLab2014, abualhaijaMachineLearningBasedApproach2019}. Thus, ISO 29148-style requirements provide a well-established and practically relevant target for studying ARI across projects and domains.

Building on this view of ARI, we investigate whether requirements can be precisely distinguished from auxiliary content across heterogeneous SRS documents to support the automatic construction of requirements datasets. While binary ARI supports scalable extraction of requirements from SRSs, we propose a ternary formulation that further divides \textit{auxiliary content} into two subcategories: \textit{(2.1) system-related auxiliary content}, which encompasses algorithms, architecture, or design, and \textit{(2.2) contextual auxiliary content}, which provides background information. This more fine-grained formulation captures richer distinctions within auxiliary content, thereby supporting the construction of more informative requirements datasets for downstream AI tasks.

Previous ARI studies indicate that BERT-style transformers outperform both traditional methods and large language models \cite{8049176, bashirRequirementNotThat2023, ivanovExtractingSoftwareRequirements2022a, wangAutomatedRequirementsDemarcation, vancanLocatingRequirementsBacklog2025}. Although emerging studies suggest that large language models achieve competitive, and in some cases higher, performance in broader requirements classification tasks \cite{el-hajjamiWhichAITechnique2023,binkhonain2025prompts}, current ARI-specific evidence supports BERT-style transformers. Accordingly, we use these models as lightweight baselines for ARI. However, ARI evaluations of both model families remain largely project- or domain-specific; while sufficient in those settings, they limit conclusions about performance on heterogeneous SRS documents and suggest that architecture alone cannot compensate for project-specific training data.

Motivated by these limitations, we address a central gap in existing ARI research: models trained on project- or domain-specific datasets may not perform reliably on heterogeneous SRS documents. To examine this gap, we construct the Automatic Requirements Identification Dataset (ARID), comprising manually collected and labeled instances from 23 heterogeneous SRSs, and use it to train and evaluate our model. To achieve our goal, we define two evaluation objectives.

\begin{textbox}{\textit{ Objective 1.}} \justifying
\noindent Identify the best-performing transformer-based model and contextualize its performance relative to a human reference point for the ARI task.
\end{textbox}

For the first objective, we ask the following question to establish a base model for ReqSeek (\textbf{Req}uirements \textbf{Seek}).

\noindent \textbf{RQ1: How well can transformer-based models identify requirements across heterogeneous SRSs?}

\noindent We evaluate RQ1 along two dimensions: \textit{(1) candidate model selection}, to identify the best-performing transformer-based model as the base for ReqSeek, given that differences in architecture, pre-training data, and optimization objectives can influence downstream performance, and \textit{(2) exploratory human-model benchmarking}, to assess manual labeling agreement and provide a human reference point for interpreting ReqSeek's performance. Having established ReqSeek, we define our second objective as follows.

\begin{textbox}{\textit{ Objective 2.}} \justifying
\noindent Evaluate ReqSeek's reliability, robustness, and generalizability for ARI, and assess its utility for supporting large-scale requirements dataset construction.
\end{textbox}

For the second objective, we ask the following research questions to further evaluate ReqSeek's reliability and utility.

\noindent \textbf{RQ2: How do increasing levels of simulated manual labeling inconsistencies affect a transformer-based model's ability to generalize to unseen requirements?}\\
Manual labeling introduces inconsistencies that may affect model learning \cite{motgerdelaencarnacionNaturalLanguageProcessing2024, liMachineLearningRequirements2022a}. We therefore conduct a post-hoc evaluation to assess the model’s sensitivity to simulated labeling noise and to provide additional evidence regarding the consistency of ARID’s labels.

\noindent \textbf{RQ3: How robust are fine-tuned transformer-based models against linguistic errors of differing severity in requirements identification?}\\
Robustness to linguistic errors becomes important when applying ReqSeek to heterogeneous and previously unseen SRS content during automated dataset construction. Therefore, we investigate whether ReqSeek tolerates minor linguistic deviations while remaining sensitive to major violations of ISO 29148-style formulation.

\noindent \textbf{RQ4: How effectively do transformer-based models identify requirements in authentic requirements SRSs?}\\
Because performance on a held-out set may reflect overfitting, we conduct a post-hoc out-of-distribution (OoD) evaluation on unseen SRSs with requirements independently labeled by their original authors to assess requirement identification beyond ARID and support the label quality of ARID.

After evaluating ReqSeek's robustness and OoD performance, we apply it to a large requirement textual dataset and ask:

\noindent \textbf{RQ5: How effectively can a transformer-based model refine a large heuristic-derived requirements corpus?}\\
We investigate whether ReqSeek can refine a large, heuristic-derived, weakly labeled requirements corpus into the Software Requirements Dataset (SwaRD) and whether the resulting dataset improves upon the original while preserving properties comparable to those of smaller, human‑curated datasets.

By addressing our evaluation objectives, we achieve our stated goal and make the following contributions to requirements engineering research:
\begin{enumerate}
\item \textbf{A transformer-based ARI model for heterogeneous SRSs:} We introduce ReqSeek, a transformer-based model for sentence-level identification of ISO~29148-style requirements across heterogeneous SRS documents to support requirements dataset construction.

\item \textbf{A manually labeled heterogeneous ARI dataset:} We construct ARID, a manually labeled dataset of requirements, system-related auxiliary, and contextual auxiliary sentences from 23 heterogeneous SRSs to support ARI training and evaluation beyond project- and domain-specific settings.

\item \textbf{A robustness and generalization evaluation protocol:} We evaluate ReqSeek beyond held-out testing through simulated labeling inconsistency, linguistic error perturbation, human-model benchmarking, human annotation and model prediction time comparison, and OoD evaluation to assess reliability, robustness, and practical applicability.

\item \textbf{A large-scale requirements dataset construction use case:} We apply ReqSeek to refine a weakly labeled corpus and construct SwaRD, a large-scale dataset of ISO~29148-style requirements from industrial SRSs, demonstrating ARI's practical utility for AI-oriented requirements dataset construction.

\item \textbf{A well-documented replication package:} We provide a replication package with publicly distributable dataset versions, annotation guidelines, model-loading scripts, evaluation scripts, experimental logs, and documentation to support reproducibility, transparency, independent verification, and reuse. Section~\ref{sec:da} presents long-term accessibility and licensing details for the replication package, ARID, and SwaRD. Appendix~\ref{apx:arid} and \ref{apx:sward} contain ARID and SwaRD dataset descriptions, metadata, column descriptions, and examples, respectively.

\end{enumerate}

\noindent \textit{Organization.} Section~\ref{sec:background} provides background; Section~\ref{sec:relatedw} reviews prior work; Section~\ref{sec:reqseek} introduces ARI with ReqSeek; Section~\ref{sec:eval} details evaluation method; results are in Section~\ref{sec:resanddisc}; implications, threats to validity, and conclusions follow in Sections~\ref{sec:implic}–\ref{sec:conc}; data availability and license info is in Section~\ref{sec:da}.

\section{Background}\label{sec:background}
This section presents fundamental concepts and the specific definitions from ISO 29148 relevant to our work.

\subsection{Transformer-Based Models}
Transformer models use self-attention to compute context-aware word representations by weighting relationships based on their surrounding context \cite{vaswaniAttentionAllYou2017}.

Nowadays, transformers are typically categorized into three families \cite{koraComprehensiveReviewTransformers2023}: encoder-only models (e.g., BERT, RoBERTa, DeBERTa) widely used for classification and understanding; decoder-only models (e.g., GPT, LLaMA, Mistral) optimized for generative tasks; and encoder–decoder models (e.g., T5, BART) optimized for sequence-to-sequence tasks such as translation or summarization.

A key factor to transformer models' success is transfer learning \cite{devlinBERTPretrainingDeep2019c, radfordImprovingLanguageUnderstandinga}, which pre-trains models on large corpora \cite{wangPreTrainedLanguageModels2023a} to learn general language features \cite{devlinBERTPretrainingDeep2019c, radfordImprovingLanguageUnderstandinga}, then fine-tunes them for specific tasks. Leveraging transformers' contextual and generalization strengths outperforms classical models trained on small domain-specific datasets \cite{prennerMakingMostSmall2022}. However, transfer learning may reduce performance on heterogeneous or unseen datasets \cite{shenGeneralizationAbilitiesFineTuned2021}, so its applicability must be carefully evaluated.

In the specific task of requirements identification, prior studies \cite{bashirRequirementNotThat2023, wangAutomatedRequirementsDemarcation, vancanLocatingRequirementsBacklog2025, koraComprehensiveReviewTransformers2023} show that encoder-only transformers consistently outperform decoder-only and encoder–decoder variants. Motivated by this empirical evidence, we focus on fine-tuning encoder-based models as lightweight baselines in our study.

Throughout this paper, we refer to encoder-only models simply as transformers, as our study focuses on them, and to decoder-only models such as ChatGPT or LLaMA as generative transformers to avoid confusion.

\subsection{ISO 29148 Requirements Specification}\label{ss:isosrs}
ISO 29148 \cite{ISOIECIEEE} provides guidance for formulating requirements that are complete, unambiguous, singular, and verifiable in systems and software engineering. Each requirement should specify a single capability, identify the subject, assign responsibility to a specific system or component, define the expected or permitted action, and state relevant conditions, object of the action, constraints, or supporting clauses within a single sentence. 

Furthermore, requirement statements should include signaling keywords to indicate their prescriptive nature. ISO 29148 specifies the modal verbs \textit{shall}, \textit{should}, \textit{may}, and \textit{will} for this purpose. However, these keywords alone are insufficient to classify a sentence as a requirement; the statement must also articulate a verifiable system obligation or provision within the context of the SRS. Generally, \textit{shall} denotes a mandatory requirement to be fulfilled under the contract; \textit{should} signifies a recommendation or preferred provision; \textit{may} conveys permission or an optional action; and \textit{will} expresses a statement of fact, futurity, or intent.

Accordingly, we define a \textit{``requirement sentence''} as a sentence-level requirement statement that satisfies the ISO 29148-style characteristics discussed above and articulates an explicit system obligation or provision. In line with ISO 29148, this definition is operationalized using the syntactic patterns below:

\begin{small}
\[
\begin{array}{ll}
\textbf{Syntax-1}: & \textit{[Subject]} \textit{[Modal Verb]} \textit{[Action]} \textit{[Constraint]} \\
\textbf{Syntax-2}: & \textit{[Condition]} \textit{[Subject]} \textit{[Modal Verb]} \textit{[Action]} \textit{[Object]} \textit{[Constraint]}
\end{array}
\]
\end{small}

For example, \textit{``The invoice system shall display pending customer invoices in ascending order of invoice due date.''} follows \textit{Syntax-1} because it specifies a subject, a modal signaling keyword, an action, and a verifiable constraint. Similarly, \textit{``At sea state 1, the radar system shall detect targets at ranges out to 100 nautical miles.''} follows \textit{Syntax-2} because it includes an explicit condition before the subject and required action. In contrast, sentences such as \textit{``Classic Shell is a discontinued computer program.''} and \textit{``The backend uses PostgreSQL.''} are not considered requirement sentences according to ISO 29148, although they are grammatically correct English sentences. The former is a factual description rather than a required system obligation or provision, while the latter describes an implementation detail rather than a required system obligation or provision.

\section{Related Work}\label{sec:relatedw}
While many studies explore Automatic Requirements Identification (ARI), a key gap persists: no existing work investigates a domain-agnostic model that automatically extracts ISO 29148-compliant requirements from heterogeneous SRSs to construct datasets to support requirements engineering research.

Early ARI studies primarily focused on traditional supervised machine learning methods. \citet{falknerIdentifyingRequirementsRequests2019} employs a Na\"ive Bayes classifier to identify requirements in request for proposal documents (RFP). Their evaluation uses 10 railway RFPs: 6 for training and 10-fold cross-validation (CV), and 4 as held-out test documents. They report 85.06\% recall and 72.04\% specificity, but provide limited detail on the expert-labeling protocol, quality-control process, inter-annotator agreement, potential human error, or the effect of labeling uncertainty on model performance. Their held-out evaluation also shows sensitivity to dataset composition when they test an RFP with a substantially different requirement/prose distribution, leaving its applicability within the intended domain open.

\citet{abualhaijaMachineLearningBasedApproach2019} applies a Random Forest classifier to identify requirements in textual specifications. They evaluate the approach on 33 industrial specifications, using a 10-fold CV for model selection and held-out specification evaluation. Their best-performing evaluation approach, which assigns false negatives twice the penalty of false positives, increases recall from 86.1\% to 95.7\%, while reducing precision from 88.1\% to 81.2\%. However, because the full industrial dataset is not publicly released, replicability is limited, and broader applicability is difficult to assess.

Relatedly, \citet{winklerAutomaticClassificationRequirements2016a} uses a convolutional neural network (CNN) to distinguish requirements from auxiliary content in automotive specifications. Their model is trained on 89 specifications from a single automotive company and evaluated with a 90/10 split, followed by evaluation on one unseen automotive specification, for which it reports 73\% precision and 89\% recall for the requirement class while tending to over-predict requirements. The authors note that applicability may be limited for documents with different vocabularies and structures.

Taken together, these studies demonstrate that early supervised ARI approaches can support requirement identification within specific industrial domains. While such domain- and setting-specific evaluations align with the studies’ intended use cases, it remains unclear whether these approaches transfer to general-purpose ARI across heterogeneous SRSs and broader requirements engineering contexts.

More recent ARI studies extend this line of work by comparing traditional machine learning pipelines with transformer-based models for requirement identification. \citet{bashirRequirementNotThat2023} evaluates different pipelines, including traditional machine learning models and transformers, for extracting requirements from five domain-specific railway industry tender documents. They use a 5-fold CV across these five tender documents and report that BERT-based classifiers achieve the highest performance, with $F_1$ scores of 82\%-87\%. In a focused encoder-based transformer setting, \citet{ivanovExtractingSoftwareRequirements2022a} fine-tunes BERT on a subset of the PURE corpus consisting of 30 SRS documents. They evaluate the model using a train/validation/test setting and report 92\% precision, 80\% recall, and an $F_1$ score of 86\% in-domain. However, performance drops to 90\% precision and 71\% recall on cross-domain documents, even after additional fine-tuning, highlighting the recall's sensitivity to domain shift.

Together, these studies provide useful evidence for transformer-based ARI within their respective evaluation settings. At the same time, results reported by \citet{bashirRequirementNotThat2023} are based on five railway tender documents, while \citet{ivanovExtractingSoftwareRequirements2022a} report reduced recall on cross-domain documents. Therefore, broader transferability for identifying requirements across heterogeneous documents remains open.

In this context, emerging research compares encoder-based transformers against generative language models. \citet{wangAutomatedRequirementsDemarcation} empirically evaluate BERT and DeBERTa against LLaMA for the task of requirements identification using the PURE dataset, consisting of 30 SRSs, and Dronology. Their evaluation combines a 5-fold CV with a 70/30--train/test. Their results indicate that DeBERTa achieved the strongest performance in their evaluation, outperforming BERT and the LLaMA model, with an $F_1$ score of 83\%. While this study provides valuable evidence on encoder and generative transformer comparisons for ARI, its conclusions remain bounded by the reused pre-labeled datasets from previous ARI studies and their limitations. Another closely related study by \citet{vancanLocatingRequirementsBacklog2025} investigates ARI in agile backlog items and compares the potential of transformer and generative transformer architectures.  They manually annotate 14 projects—half from a single company—for requirements identification and classification, and evaluate their models using a 7-fold project-level CV. Their results also demonstrate that RoBERTa achieves the strongest performance among the evaluated models, outperforming Llama 3, Mistral 7B, and GPT-4, with an average $F_1$ score of 0.82 for requirement identification. 

Both studies provide useful evidence that encoder-based transformer models may be well-suited for ARI-style identification tasks. However, the authors acknowledge limits to broader generalizability for identifying requirements in heterogeneous documents: \citet{wangAutomatedRequirementsDemarcation} reports that generalizability to other datasets and identification criteria remains uncertain, while \citet{vancanLocatingRequirementsBacklog2025} reports the absence of cross-domain and out-of-distribution evaluation.

Further, the comparative suitability of encoder-based and generative models for general-purpose ARI should be viewed as preliminary and warrant further study, particularly given broader requirements engineering studies reporting competitive zero-shot and few-shot results for generative transformers. For example, \citet{el-hajjamiWhichAITechnique2023} reports that no single technique among SVM, LSTM, GPT-3.5, and GPT-4 performs best across all requirement classes, while \citet{binkhonain2025prompts} reports that prompt-based generative transformer models can reach performance comparable to or exceeding a fine-tuned encoder-based model in several requirements classification tasks. Our study complements this line of work by evaluating lightweight encoder-based models for ARI-specific generalization across heterogeneous SRSs, with attention to their potential as less resource-intensive alternatives.

A structured empirical comparison of the ARI studies discussed in this section is provided in Table~\ref{tab:empirical-comparison} to situate ReqSeek directly against current ARI approaches.

Building on previous research and addressing recurring limitations in prior work -- specifically (i) the lack of cross-domain training datasets and (ii) the absence of generalization evaluation beyond the training environment -- we fine-tune ReqSeek on a requirements dataset from 23 heterogeneous SRSs to support its domain-agnosticism for ARI. We evaluate ReqSeek through human-model benchmarking and on unseen SRSs with requirements independently labeled by their original authors to assess domain-agnostic requirement identification. Unlike prior studies, we further characterize whether ReqSeek's identified requirements align with the textual and semantic properties of human-curated requirements datasets. Additionally, we analyze manual labeling inconsistencies and linguistic error sensitivity to provide further evidence of ReqSeek's robustness.

\section{Automatic Requirements Identification with ReqSeek}\label{sec:reqseek}
\begin{figure}[ht]
    \centering
    \includegraphics[width=1\linewidth]{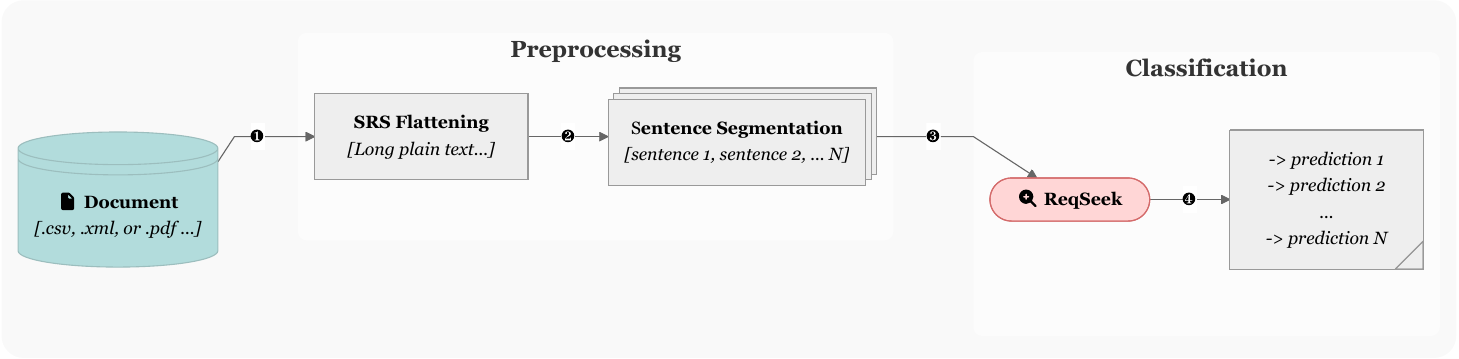}
    \caption{Automated Requirements Identification (ARI) with ReqSeek.}
    \Description{This figure illustrates the three-step pipeline for automatically identifying requirements in textual documents. First, (1) input documents (such as CSV, XML, or PDF files) are parsed and flattened into raw text. Next, (2) the flattened text is segmented into individual sentences. Finally, (3) each sentence is passed to ReqSeek for ternary classification as a requirement, system-related auxiliary, or contextual auxiliary.}
    \label{fig:reqseek}
\end{figure}

ReqSeek (\textbf{Req}uirements \textbf{Seek}) is a fine-tuned transformer model designed to automatically identify ISO 29148-style requirements within heterogeneous SRS documents. Figure~\ref{fig:reqseek} presents the workflow of ReqSeek. The process consists of three steps: \textit{(1) flattening the SRS into plain text}, \textit{(2) segmenting the text into sentence-level units}, and \textit{(3) inputting each sentence-level unit} into ReqSeek for classification.

Sentence-level units are derived from the flattened SRS using punctuation-based sentence segmentation, where sentence-ending punctuation marks (e.g., ``.'', ``!'', and ``?'') are used to identify candidate sentence boundaries. These units are subsequently interpreted according to the requirement-sentence definition in Section~\ref{ss:isosrs}. Each sentence-level unit is treated as a candidate unit for classification. Accordingly, a requirement sentence may contain multiple clauses, such as conditions, constraints, or temporal qualifiers, provided they contribute to a single required capability, characteristic, or constraint. Conversely, a sentence that expresses several independent obligations may constitute more than one functionality in a given requirement, which should ideally be decomposed into simpler, complete, singular, and verifiable requirement sentences according to ISO 29148.

Consequently, ReqSeek classifies each sentence independently as a requirement, system-related, or contextual auxiliary content, and does not merge adjacent sentences into a single multi-sentence requirement. This approach is consistent with prior requirements engineering studies \cite{abualhaijaMachineLearningBasedApproach2019} and reflects the textual properties of small, human-curated requirements datasets \cite{limaSoftwareEngineeringRepositories2019, Frieden84Nlp4rereqevalV20}. 

Nevertheless, requirements in real-world SRS documents do not always strictly follow ISO 29148-style formulation. Complex, compound, or distributed requirements may arise in practice. For example, a single functionality may be described across multiple adjacent requirement sentences, each containing its own subject, action, condition, or constraint. Conversely, multiple functionalities may be expressed within a single sentence, often through conjunctions such as ``and'' or ``or''. Such cases may still be identified when the requirement statements remain reasonably aligned with the requirement-sentence formulation described in Section~\ref{ss:isosrs}. However, the expected subject, modal verb, action, condition, or constraint may also be implicit or structurally displaced, or the requirement statement may be intermixed with substantial contextual, explanatory, or design-related information that obscures the underlying system obligation or provision.

In such cases, sentence-level classification may only partially capture the requirement, classify related fragments separately, or fail to identify requirements with major deviations from the structure of a self-contained requirement sentence. This limitation is particularly relevant for dataset construction, in which including highly ambiguous, structurally incomplete, or substantially non-compliant requirement fragments may reduce dataset quality and affect downstream uses of the curated requirements.

To partially account for such variation, ARID construction includes practical cases observed in SRS documents, such as different modal verbs, compound requirement statements, and non-requirement instances with requirement-like signaling keywords or syntactic structures. ReqSeek is further evaluated under minor and major textual transformations to assess its robustness to practical deviations from ideal requirement formulation and to better understand the extent to which such deviations affect requirements identification. The ARID dataset and its construction are described in the subsequent section.

\subsection{Training Dataset}
Below, we describe our process for curating a suitable dataset, the Automatic Requirements Identification Dataset (ARID), to train ReqSeek for Automatic Requirements Identification (ARI) using ISO 29148-style requirements.

\subsubsection{SRS Selection for Dataset}
To identify publicly available sources for curating a suitable ARI dataset, we conducted targeted searches across multiple search engines using the specific query keywords listed in Table \ref{table:searchandkeywords}. Additionally, the following criteria were applied after identifying each source:
\begin{enumerate}
\item The source must originate from a real-world software project.
\item The source must contain ISO 29148-style requirements.
\end{enumerate}

\begin{table}[ht]
\centering
\caption{Search Engines and Keywords for Sourcing Requirements}
\label{table:searchandkeywords}
\begin{tblr}{
  width = \linewidth,
  colspec = {Q[140]Q[425]Q[373]},
  row{1} = {c},
  cell{2}{2} = {r=2}{},
  cell{2}{3} = {r=2}{},
  cell{5}{2} = {t},
  vlines,
  hline{1-2,4-6} = {-}{},
  hline{3} = {1}{},
}
\textbf{Search Engine} & \textbf{Queries Keywords} & \textbf{Filter}\\
{\textbf{Google }\\\textbf{Dataset Search}} & {Software Requirements, Functional Requirements, \\Non-functional Requirements, Requirements Traceability} & Results with the actual dataset were considered.\\
\textbf{Zenodo} &  & \\
\textbf{Google Scholar} & Software Requirements Dataset & Datasets linked to research papers\\
\textbf{Google Search} & Software Requirements Specification & Filetype: pdf, docx, doc, xlsx
\end{tblr}
\end{table}

This process identified 23 publicly available requirements sources across multiple domains. The identified sources originate from academic, industrial, and community benchmark settings, including industrial or multi-organization research initiatives such as Gaia-X~\cite{GaiaX}, ACTIVAGE~\cite{pilarsala20215666974}, and OPENCOSS~\cite{OPENCOSS}, as well as publicly curated academic benchmark corpora and RE datasets such as Dronology~\cite{8444851}, FN-RE~\cite{waad_alhoshan_2018_1291660}, MobSTr, and the Online Feedback Crowd Dataset~\cite{van_vliet_martijn_2020_3626185}. Representative application domains include healthcare, aerospace, automotive safety-critical systems, cloud infrastructure, cyber-physical systems, enterprise software, Internet of Things (IoT) for smart home devices, telecommunications, and mobile applications. Interested readers can find detailed metadata and reference information for all source datasets and SRSs in Appendix~\ref{apx:arid}.

\subsubsection{Training Instances Selection and Labeling}
To extract requirements, we flattened the 23 SRS documents, removed HTML tags and table formatting, and split the text into sentences. The resulting sentences were manually labeled by the first author into three classes: \textit{(1) requirement}, \textit{(2) system-related}, or \textit{(3) contextual auxiliary}. Before labeling, the four authors collaboratively reviewed examples and discussed class boundaries to ensure consistency. Following the empirical study by \citet{DifferentResearchersDifferent}, we employed a single labeler with sufficient in-domain experience. Their findings show that for simple tasks with defined class boundaries, a single coder can be sufficient and provide higher internal validity, which aligns with our setup.

The distinction between the three labels was operationalized through decision rules based on intent and verifiability. Requirements express verifiable system obligations, whereas system-related auxiliary and contextual auxiliary content provide implementation-related information and document context, respectively. For example, ``The system shall use TCP/IP for communication'' was labeled as a requirement because it specifies an obligation, ``Communication uses TCP/IP'' as system-related auxiliary because it describes an implementation choice, and ``The following section describes the communication architecture'' as contextual auxiliary because it provides document context. Borderline cases were resolved by considering whether a statement specifies required system behavior rather than a human decision, implementation instruction, or document-level explanation. The complete annotation guideline, decision rules, and additional examples are available in the replication package\footnote{For transparency and reproducibility, the complete annotation guide used during ARID labeling is included in the replication repository under \texttt{evaluations/RQ1/human\_annotated\_data/Annotation\_Guide.docx}.}.

To address potential labeling subjectivity and support labeling quality, we conducted post-hoc validation through multiple analyses. As part of the \textit{exploratory human-model benchmark} dimension in RQ1, we assess inter-rater agreement on a sample using independent annotators. RQ2 examines model sensitivity to manual labeling inconsistencies, and RQ4 evaluates the model on unseen SRSs with requirements independently labeled by their original authors. Further, we provide a publicly distributable version of ARID in the replication package  (see Section~\ref{sec:da}), allowing readers to inspect its annotations and independently assess the quality of the available instances.

During curation, we observed that many requirements in the selected SRSs use the ``\textit{must}'' keyword, which is not recommended by ISO 29148. To reflect real-world variation, we retained such instances in the dataset and fine-tuning process, though they can be filtered if necessary.

A balanced number of requirements with each of the four signaling keywords was selected. To prevent models from relying solely on these keywords as shortcuts for requirements identification, balanced training instances were also drawn from the two auxiliary classes, including sentences containing signaling keywords. Since contextual auxiliaries are neither requirements nor technical content, we further augmented this class with sentences from the Wikipedia dataset available on Hugging Face \cite{LegacydatasetsWikipediaMain2024}, to provide general domain knowledge beyond what was present in the SRSs. This composition ensures the model can distinguish requirements from non-requirements while minimizing false positives.

\subsubsection{ARID Dataset Composition}\label{ssec:train_eval_set}
The complete ARID research dataset used in this study consists of 2,396 instances: 1,050 requirements, 746 system-related auxiliaries, and 600 contextual auxiliaries. Details on dataset availability and licensing are provided in Section~\ref{sec:da}, while the training and evaluation splits are described in Section \ref{ssec:evm}.

\subsection{ReqSeek Fine-tuning Setup}
Here, we describe the selected base models, fine-tuning hyperparameters, and hardware and software configuration used in our study.

\subsubsection{Selecting Models for Fine-Tuning}
For fine-tuning, we opt for open-source models to support continual learning, advancing collaboration, and promoting transparency. Additionally, open-source models ensure local usability, which is essential for organizations that prefer to deploy models for privacy and control. Smaller supervised models like RoBERTa are preferred over large LLMs (e.g., LLaMA, Gemma) due to lower overhead, easier deployment, and better classification performance \cite{yuOpenClosedSmall2023}. This preference is further evidenced by \citet{wangAutomatedRequirementsDemarcation} and \citet{vancanLocatingRequirementsBacklog2025} in the case of ARI.

To establish an initial ARI benchmark, we first fine-tuned \mbox{all-MiniLM-L6-v2}~\cite{SentencetransformersAllMiniLML6v2Hugging2024, HuggingFace}, a lightweight model optimized for sentence-level representation learning. This choice aligns with our sentence-level formulation of ARI, as discussed in Section~\ref{sec:reqseek}. We then fine-tuned and compared BERT-base-cased~\cite{GooglebertBertbasecasedHugging2024, HuggingFace}, RoBERTa-base~\cite{FacebookAIRobertabaseHugging2024, HuggingFace}, and GPT-2~\cite{OpenaicommunityGpt2Hugging, HuggingFace}. Table~\ref{tab:plm-comparison} summarizes their architectural differences and selection rationale.

\begin{table}[h]
\centering
\caption{Characteristics of the Selected Pre-trained Transformer Models for ARI Fine-tuning in this Study}
\label{tab:plm-comparison}

\setlength{\tabcolsep}{6pt}

\resizebox{\textwidth}{!}{%
\begin{tabular}{p{2.3cm} p{2.3cm} p{1.5cm} p{2.0cm} p{2.2cm} p{0.3cm} p{3.5cm}}
\toprule

\multicolumn{1}{c}{\textbf{Model}} &
\multicolumn{1}{l}{\textbf{Architecture}} &
\multicolumn{1}{l}{\textbf{Tokenization}} &
\multicolumn{1}{l}{\textbf{Training Data}} &
\multicolumn{1}{l}{\textbf{Optimization}} &
\multicolumn{1}{c}{\textbf{Size}} &
\multicolumn{1}{c}{\textbf{Rationale}} \\

\midrule

all-MiniLM-L6-v2 &
Encoder-only &
WordPiece &
\mbox{1B sentence} pairs from multiple datasets &
Contrastive sentence-level learning &
22.7M &
Lightweight sentence-level baseline suitable for initial benchmarking. \\

BERT-base-cased &
Encoder-only &
WordPiece &
BooksCorpus + Wikipedia &
Masked language modeling + next-sentence prediction &
110M &
\citet{ivanovExtractingSoftwareRequirements2022a} and \citet{bashirRequirementNotThat2023} report strong ARI performance with this model. \\

RoBERTa-base &
Encoder-only &
\mbox{Byte Pair} Encoding &
BookCorpus, CC-News, OpenWebText &
\mbox{Dynamic masked} \mbox{language modeling} &
125M &
\citet{wangAutomatedRequirementsDemarcation} and \citet{vancanLocatingRequirementsBacklog2025} report strong ARI performance with this model. \\

GPT-2 &
Decoder-only &
\mbox{Byte Pair} Encoding &
WebText &
Autoregressive \mbox{language modeling} &
117M &
Included to test whether ARID benefits decoder-only architectures reported to be less suitable for ARI by \cite{vancanLocatingRequirementsBacklog2025}. \\

\bottomrule
\end{tabular}%
}
\end{table}

\subsubsection{Fine-tuning Hyperparameters}
Throughout our experiments, we used Hugging Face Transformers v4.42.3~\cite{HuggingFace} with \texttt{TensorFlow v2.15.0}, \texttt{Keras v2.15.0}, and \texttt{PyTorch v2.3.1+cu121} sequence-classification models. To ensure a controlled comparison, all candidate models and folds, as well as the selected best-performing model, used the same explicitly specified fine-tuning setup: 30 epochs with no early stopping, batch size 16, initial learning rate $2 \times 10^{-5}$, and zero warm-up steps. Other optimizer parameters not reported here, such as weight decay, were left unchanged and inherited from the corresponding Hugging Face Transformers API defaults.

This setup follows common encoder-only fine-tuning practice and is consistent with prior RE work. In particular, 
our chosen hyperparameters are consistent
with the best-performing configurations reported by \citet{heyNoRBERTTransferLearning2020}. They used a learning rate of $2 \times 10^{-5}$, no warm-up phase, and batch size 16 for encoder-based requirements classification, reporting this setup among their best-performing configurations with 10--32 epochs. Following the recommendations from \citet{hoffmannTrainingComputeOptimalLarge2022a}, we also kept the learning rate small to facilitate stable convergence and reduce the risk of divergence during fine-tuning.

\subsubsection{Fine-tuning Hardware}

All fine-tuning experiments were conducted on a single \texttt{NVIDIA GeForce RTX 4090 GPU} with \texttt{24GB} of memory, running \texttt{Ubuntu 22.04.5 LTS} and \texttt{Python 3.10.12}.

\section{Evaluation Method}\label{sec:eval}
This section presents the evaluation method and experimental setup for identifying the best-performing transformer model for ReqSeek and assessing its performance.

\subsection{Study Design} \textbf{\textit{Our goal is to develop a domain-agnostic model for Automatic Requirements Identification (ARI) that identifies requirements formulated in compliance with ISO 29148 and supports the systematic construction of large-scale requirements datasets for AI applications in requirements engineering research.}} To achieve our objective, we employ a multi-method evaluation design through two evaluation objectives and five research questions. For the \textit{candidate model selection} dimension of RQ1, as well as for RQ2 and RQ3, we use controlled experiments. These experiments systematically vary model- or data-related conditions to observe their effects on model behavior, thereby supporting internally valid comparisons \cite{diesteQuantitativeDeterminationRelationship2011, balakrishnanCausalBenchmarkingBiasin2021}.

RQ1 further evaluates an \textit{exploratory human-model benchmarking}, which provides a human reference point for interpreting ReqSeek's performance and effort estimation, and assessing the consistency of ARID labels with independent human judgments. RQ4 addresses external generalization through an out-of-distribution (OoD) evaluation on previously unseen SRSs, assessing ReqSeek's robustness beyond the training distribution. Finally, RQ5 performs use-case validation by constructing a dataset identified by ReqSeek and characterizing its textual and semantic properties.

\begin{figure}[ht]
    \centering
    \includegraphics[width=1\linewidth]{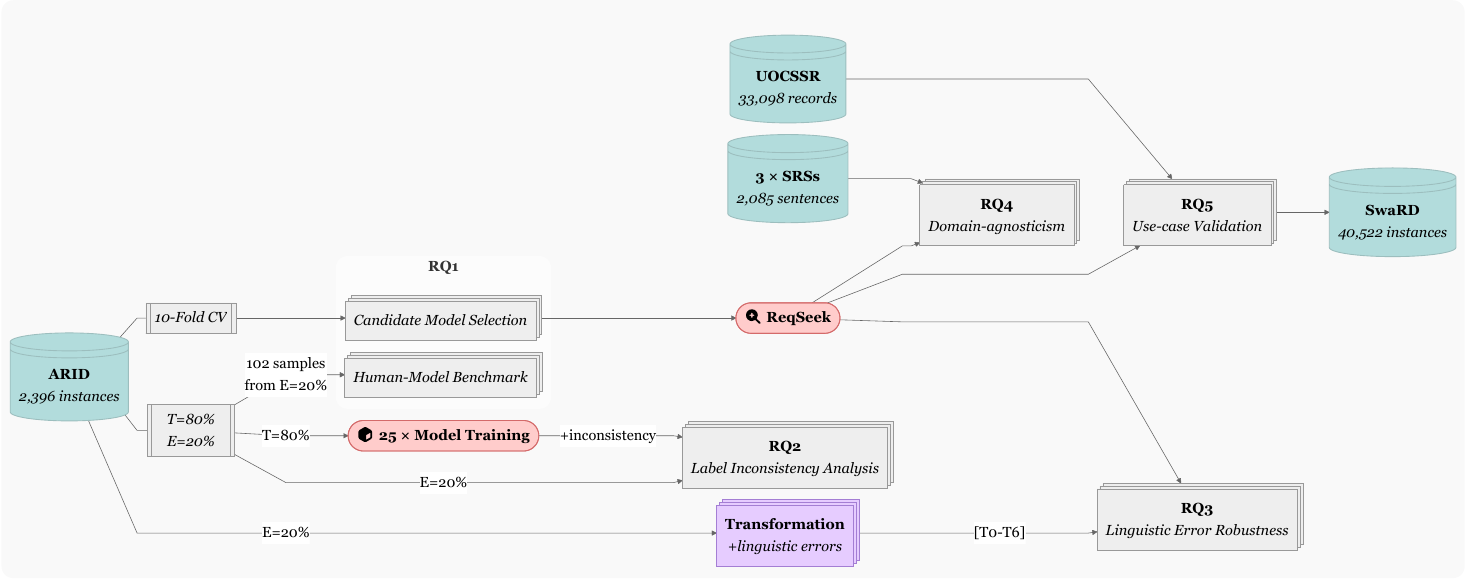}
    \caption{Schematic Overview Linking Study Design's Main Components. Abbreviations: T = Training Set; E = Evaluation Set.}
    \Description{This schematic provides a high-level overview of the core components in the study design and their relationships. It visually connects the main elements and illustrates how they interact throughout the experimental pipeline.}
    \label{fig:method}
\end{figure}

This multi-method approach evaluates ReqSeek from multiple perspectives, balancing internal validity with practical relevance. Fig.~\ref{fig:method} summarizes the different parts of the design, which are described in the following subsections in more detail.

\subsection{Tasks}
We address our research questions through the following tasks.

\subsubsection{RQ1 Model Selection and Human Benchmarking}\label{ssec:mb}\mbox{}

\noindent \textit{Candidate Model Selection Dimension.} We follow the foundational work on robust classifier comparison and model selection using cross-validation by \cite{kohaviStudyCrossvalidationBootstrap1995}. We fine-tune and compare multiple transformer-based models using stratified cross-validation to estimate their generalization performance and select the best-performing architecture. The selected model is then re-initialized from its pre-trained weights and fine-tuned from scratch to obtain the ReqSeek model, which is used for deployment and in all subsequent experiments. This two-stage setup separates model selection from downstream evaluation, thereby preventing information leakage.

\noindent \textit{Exploratory Human-Model Benchmarking Dimension.} We evaluate ReqSeek against independent human annotations on the same ARI labeling task. 

We recruited annotators with experience in software engineering and requirements artifacts. To calibrate their understanding of the ARI task, they were provided with an annotation guideline covering the class definitions, decision rules, and examples. Before annotation, they attended a 15-minute online onboarding session introducing the study objective, datasets, task procedure, and a brief ISO 29148 knowledge refresher. They then familiarized themselves with the labeling tool for approximately five minutes, asked questions, and labeled practice examples. Annotators also received an evaluation package that included a link to the online labeling tool, an annotation guide, and an anonymous background questionnaire, and were instructed to submit their responses anonymously.

We first perform inter-rater agreement analyses to assess the consistency of ARID's labeling process and the alignment of ReqSeek's predictions with independent human annotations. Subsequently, we compare ReqSeek's performance with that of human annotators to provide a human reference point for interpreting model performance and comparing human annotation and model prediction times in the context of dataset construction.

\subsubsection{RQ2 Labeling Inconsistency Analysis} We evaluate the impact of simulated labeling inconsistencies (i.e., randomly flipping the original class labels) on the base model selected for ReqSeek. The selected baseline model's pre-trained weights are re-initialized and fine-tuned on versions of the training set with increasing levels of labeling inconsistency, while the evaluation set remains unchanged. Model performance is systematically measured to assess how it degrades as mislabeling in training data increases.

\subsubsection{RQ3 Linguistic Errors Robustness Evaluation} We apply minor transformations (e.g., word order errors or missing details) and major transformations (e.g., ISO 29148-style violations or missing signaling keywords) to simulate linguistic errors in human-written requirements and evaluate whether ReqSeek tolerates minor linguistic errors while remaining sensitive to major deviations. With minor transformations, we aim to simulate acceptable linguistic variation that does not severely affect ISO 29148-style (see Section~\ref{ss:isosrs}), whereas major transformations simulate severe violations of the standard.

\subsubsection{RQ4 Domain-Agnosticism Evaluation} 
To assess ReqSeek's domain-agnostic requirement identification, we perform a post-hoc OoD evaluation using unseen SRSs with requirements independently labeled by their original authors.

\subsubsection{RQ5 Use-Case Validation} 
We apply ReqSeek to a large, weakly labeled requirements corpus previously constructed using heuristic methods. The objective is to use ReqSeek to classify candidate instances as requirements, system‑related auxiliaries, or contextual auxiliaries, thereby creating a new dataset, which we name the Software Requirements Dataset (SwaRD).

We then examine, at the corpus level, whether the requirements identified by ReqSeek differ in their textual and semantic properties from those previously obtained through heuristic extraction and whether they align with established human-curated requirements datasets. Furthermore, we assess whether SwaRD achieves substantially greater scale and diversity while maintaining comparable quality. This analysis focuses on the characteristics of the resulting dataset rather than on direct comparison of the identification methods.

\subsection{Training and Evaluation Materials}\label{ssec:evm}
This section describes the training and evaluation materials used for each research question. 

\subsubsection{RQ1 Evaluation Material}\mbox{}

\noindent \textit{Candidate Model Selection Dimension.} We create a stratified 10-fold cross-validation on the ARID dataset (for dataset creation details, see Section \ref{ssec:train_eval_set}), in which each split consists of 80\% training data and 20\% validation data. Next, we build a fixed-random single-split with 80\% training and 20\% evaluation data from ARID to fine-tune and evaluate ReqSeek.

\noindent \textit{Exploratory Human-Model Benchmarking.} We draw a stratified sample of 102 instances (for sample-size estimation, see Section~\ref{ssec:sample_size}) from ARID's fixed 20\% evaluation split, which remained unseen by ReqSeek during training. For this subset, the evaluation material consists of independent human annotations, consensus labels derived from these annotations, first-author ARID labels, ReqSeek predictions, and per-instance, per-annotator annotation-time measurements.

\subsubsection{RQ2 Evaluation Material} We use the same fixed 80/20--train/test split. On the 80\% training portion, we inject labeling inconsistencies at levels of 0\% (original labels, no inconsistency is added), 10\%, 20\%, 30\%, and 40\%, resulting in four differently corrupted versions of the training set, while keeping the evaluation set clean.

\subsubsection{RQ3 Evaluation Material} We systematically applied six textual transformations to the created fixed 20\% evaluation data in RQ1. Table~\ref{tab:transformations} details these transformations. Each transformation resulted in a new, unseen evaluation set, totaling six.

\begin{table}[ht]
\centering
\caption{Textual Transformations Simulating Minor and Major Linguistic Errors}
\label{tab:transformations}
\begin{tblr}{
  width = \linewidth,
  colspec = {Q[38]Q[87]Q[356]Q[458]},
  row{1} = {c},
  cell{2}{1} = {c},
  cell{2}{2} = {c},
  cell{3}{1} = {c},
  cell{3}{2} = {c},
  cell{4}{1} = {c},
  cell{4}{2} = {c},
  cell{5}{1} = {c},
  cell{5}{2} = {c},
  cell{6}{1} = {c},
  cell{6}{2} = {c},
  cell{7}{1} = {c},
  cell{7}{2} = {c},
  hline{1,8} = {-}{0.08em},
  hline{2} = {-}{},
}
\textbf{ID} & \textbf{Severity} & \textbf{Transformation} & \textbf{Description}\\
T1 & Minor & Add Minor Details & Random extra details are added\\
T2 & Minor & Re-order Words & Random words positions are swapped\\
T3 & Minor & Remove Minor Details & Random details are removed\\
T4 & Major & {Signaling\\Keywords Removed} & ISO 29148 signaling keywords are removed\\
T5 & Major & {ISO Syntax Violation\\(+Signaling Keywords)} & T1 through T3 are simultaneously applied\\
T6 & Major & {ISO Syntax Violation\\(-Signaling Keywords)} & T1 through T4 are simultaneously applied
\end{tblr}
\end{table}

To realize T1, we add minor details in the form of short descriptive phrases (e.g., ``to provide an effective solution'') to simulate requirements that are overly detailed. T2 reorders words to evaluate tolerance for grammatical errors, e.g., misplaced conditions or subjects. T3 removes words, reflecting that some information is missing. In contrast, T4--T6 represent major linguistic errors severely violating the ISO 29148-style. T4 removes all signaling keywords. T5 distorts the ISO 29148-style requirement while keeping the signaling keywords. T6 distorts both syntax and keywords, simulating a complete compliance violation.

\subsubsection{RQ4 Evaluation Material} We use the following criteria to select real-world SRSs.
\begin{itemize}
    \item Must be a real software project 
    \item Must contain ISO 29148-style requirements
    \item Must contain a dedicated requirements section in which each requirement is explicitly listed with a unique ID. These requirement IDs were used solely to derive ground-truth labels after sentence-level parsing, enabling the computation of the performance metrics reported for the OoD evaluation; they were not provided to ReqSeek as model input.\footnote{In practical deployment, users can apply ReqSeek directly to sentence-level text without manually identifying requirement sections or assigning requirement IDs beforehand.}
\end{itemize}

The three selected SRSs matching our selection criterion are the Restaurant Menu \& Ordering System (RMOS) \cite{senkyrSoftwareRequirementsData2023}, the Nodes Portal Toolkit (NTP) \cite{senkyrSoftwareRequirementsData2023}, and 
{K\"unstliche Intelligenz in die berufliche Bildung bringen (KIB\textsuperscript{3})}. KIB\textsuperscript{3} is an AI educational initiative based in Baden-W\"urttemberg, Germany, focused on developing AI tools to support students. It is currently proprietary. We have made it open source as part of this work.

\subsubsection{RQ5 Evaluation Material}\label{ssec:rq5evm} We use the Unified Open and Closed Source Software Requirements (UOCSSR) dataset as the source corpus for this study. UOCSSR was curated using heuristic methods in our previous study; refer to \citet{senger2022unified} for details on its design and construction.

The original UOCSSR dataset contains 43,777 records from 513 SRSs and requirements datasets across eight requirement types. It includes both publicly available and proprietary records. In this work, we use only the publicly available subset, comprising 33,098 records. Of these, 12,080 records were identified as ISO 29148-style requirements by the heuristic method.

Because the heuristic extraction approach for identifying ISO 29148‑style requirements was not validated against any manually labeled data (neither a sample nor the full set), its performance is unknown. Therefore, we treat the resulting labels as weak labels for this study. We use all 33,098 public records as source text for sentence-level requirement identification with ReqSeek. The corpus is flattened and segmented into candidate requirement sentences, as defined in Section \ref{ss:isosrs}, yielding 40,522 sentence-level records.

To this end, we create two separate datasets from UOCSSR. The first is the heuristic-derived ISO 29148-style requirement subset, consisting of 12,080 records; we refer to this dataset as UOCSSR-ISO. The second is the 40,522 sentence-level corpus, classified by ReqSeek into requirements, system-related auxiliaries, and contextual auxiliaries; we refer to this dataset as the Software Requirements Dataset (SwaRD). 

In addition, for the human‑curated datasets, we use PROMISE‑exp \cite{limaSoftwareEngineeringRepositories2019} and NLP4RE‑ReqEval \cite{sallam_abualhaija_2021_4471411, Frieden84Nlp4rereqevalV20}, which in this study we simply refer to as PROMISE and NLP4RE, respectively.

\subsection{Analysis Procedure}\label{ssec:analp}
Except for RQ5, we compute and report precision, recall, and $F_1$ to evaluate model performance under conditions aligned with each objective, as defined in the following equations. These metrics are reported under the evaluation conditions associated with each research question. In the equations and overall our study, TP, FP, TN, and FN denote true positives, false positives, true negatives, and false negatives, respectively.\\

\begin{minipage}{0.30\textwidth}
\begin{equation}
\mathrm{Precision} = \frac{TP}{TP + FP}
\label{eq:precision}
\end{equation}
\end{minipage}
\hfill
\begin{minipage}{0.30\textwidth}
\begin{equation}
\mathrm{Recall} = \frac{TP}{TP + FN}
\label{eq:recall}
\end{equation}
\end{minipage}
\hfill
\begin{minipage}{0.30\textwidth}
\begin{equation}
F_1 = 2 \cdot \frac{\mathrm{P}\cdot\mathrm{R}}{\mathrm{P}+\mathrm{R}}
\label{eq:f1}
\end{equation}
\end{minipage}\\

Following the view of \citet{berryEmpiricalEvaluationTools2021} that evaluation criteria should reflect the intended task context, the relative importance of precision and recall depends on the evaluation objective. For requirement identification tasks, Berry emphasizes recall-oriented evaluation, especially when the objective is to discover as many relevant instances as possible. We agree with this position for discovery-oriented settings. In this work, however, ReqSeek is intended to support the construction of high-quality requirements datasets. In this setting, precision becomes central because false positives introduce incorrect instances and directly compromise dataset quality. Recall remains important because it reflects coverage; missed requirements reduce the dataset size rather than contaminating it. Accordingly, we emphasize precision in evaluating ReqSeek, while also reporting recall as a measure of coverage. Further, $F1$ is reported for completeness and to facilitate easier comparison with prior work using a single metric.

For RQ1, RQ2, and RQ3, we use statistical tests with a significance level of $\alpha = 0.05$ and report 95\% confidence intervals for effect size estimates when applicable.

\subsubsection{RQ1 Analysis Procedure} For both evaluation dimensions, in addition to the standard performance metrics defined above, we compute the task-weighted $F_\beta$ metric following \citet{berryEmpiricalEvaluationTools2021}. Unlike $F_1$, $F_\beta$ allows precision and recall to be weighted according to the evaluation objective when computing an aggregate performance measure. In our study, $F_\beta$ is weighted for precision. $F_\beta$ is computed as follows:

\begin{equation}
F_\beta = \frac{(1 + \beta^2)\,\text{TP}}{(1 + \beta^2)\,\text{TP} + \text{FP} + \beta^2\,\text{FN}}
\end{equation}\\

A value of $\beta < 1$ places greater weight on precision, whereas $\beta > 1$ places greater weight on recall. We empirically derive the parameter $\beta$ in our study and detail its computation in the \textit{exploratory human-model benchmark} dimension.

\noindent \textit{Candidate Model Selection Dimension.} We evaluate models via 10-fold cross-validation and summarize fold-level precision scores using the mean ($\hat{x}$) and standard deviation ($s$). To select the base model, we compare fold‑level precision scores of fine‑tuned candidates using the Friedman test and quantify differences with Kendall’s 
$W$ effect size, and test the hypothesis below.

\begin{textbox}{\textit{ Hypothesis.}} \justifying
\noindent $H_{0,1}$: There is no significant difference in precision across the 10 folds among the fine-tuned candidate models.

\noindent $H_{a,1}$: At least one fine-tuned candidate model differs significantly in precision across the 10 folds.
\end{textbox}

We then apply Nemenyi's post-hoc test to identify significant differences among candidate models. The selected model is re‑initialized from its pre‑trained weights, retrained on the fixed training set to produce ReqSeek, and evaluated on the corresponding fixed evaluation set. For ReqSeek, we report per-class precision, recall, $F_1$, and $F_\beta$.

\noindent \textit{Exploratory Human-Model Benchmark Dimension.} Annotators independently label each sampled instance. The labeling tool internally records per-instance annotation time without informing the annotators, to reduce the potential influence of timing awareness on labeling behavior. We then derive consensus labels by majority vote and use them as the reference for analysis.

We assess inter-rater agreement among the annotators using Fleiss' $\kappa$. We then compute Cohen's $\kappa$ between the consensus labels and each comparison source: the independent annotators, the first-author ARID labels, and ReqSeek's predictions. This setup lets us first measure inter-rater agreement among annotators, then compare each annotator's and the first-author ARID labels against the consensus reference to assess both individual annotator consistency and the consistency of ARID labels with independent human judgments, and finally assess ReqSeek's alignment with human labeling behavior.

Next, to provide a human reference point for ReqSeek's performance, we compare ReqSeek, the first-author labels, and each independent annotator against the consensus labels using precision, recall, $F_1$, and $F_\beta$. We also report total human annotation and model prediction time in minutes. In addition, following the timing-based interpretation proposed by \citet{berryEmpiricalEvaluationTools2021}, we report the mean time to find true positives and the mean time to decide false positives in seconds. The first-author annotation time is reported as unavailable because it was not recorded during labeling. ReqSeek timing reflects the total prediction time for an instance, including preprocessing and model inference, and is interpreted as hardware-specific. Inference was run on a Mac with an Apple M4 chip as a commercially available platform.

To empirically determine the parameter $\beta$ for $F_\beta$, we divide each annotator's mean time to find true positives by their mean time to decide false positives and average the resulting ratios across the three annotators.

Further, we report prevalence and output fraction. Prevalence denotes the proportion of instances belonging to the target class in the reference labels, whereas output fraction denotes the proportion of instances assigned to the target class by a comparison source. These measures are calculated as follows:

\begin{center}
\begin{minipage}{0.35\textwidth}
\centering
\begin{equation}
\mathrm{Prevalence} = \frac{TP + FN}{N}
\label{eq:prevalence}
\end{equation}
\end{minipage}
\hspace{0.05\textwidth}
\begin{minipage}{0.35\textwidth}
\centering
\begin{equation}
\mathrm{Output\ Fraction} = \frac{TP + FP}{N}
\label{eq:output_fraction}
\end{equation}
\end{minipage}
\end{center}

Here, $N$ represents the total number of evaluated instances. The sum $TP + FN$ indicates the number of reference instances belonging to the target class, while $TP + FP$ represents the number of instances assigned to the target class by a comparison source.

\subsubsection{RQ2 Analysis Procedure} For each inconsistency level, we conduct five fine-tuning runs with the identified best-performing model, re-initializing it from its pre-trained weights with a unique random seed per run.

Since the labels at 0\% inconsistency must remain unchanged, we use different random seeds across five fine-tuning runs to control for randomness. For the rest of the simulated inconsistency levels, we retrain RoBERTa-base five times, each time flipping a different random subset of labels only within the training set. We record precision scores for each run, reflecting our focus on dataset quality, to collect performance samples at each inconsistency level. This approach captures performance variability due to label inconsistencies rather than factors such as data shuffling or model initialization. Further, we report sample size ($n$), mean ($\hat{x}$), and standard deviation ($s$). To systematically assess performance degradation across label inconsistencies, we grouped the scores by inconsistency level. Before conducting one-way ANOVA, we assessed its assumptions. Normality was evaluated using Shapiro-Wilk tests for each labeling inconsistency level, and homogeneity of variances was evaluated using Levene's test. We then performed a one-way ANOVA and quantified its magnitude using Partial Eta Squared ($\eta_p^2$) to test the following hypothesis.

\begin{textbox}{\textit{ Hypothesis.}} \justifying
\noindent $H_{0,2}$: There is no significant difference in precision across inconsistency levels.

\noindent $H_{a,2}$: At least one inconsistency level has a statistically significant difference in precision.
\end{textbox}

Then, we apply Tukey's honestly significant difference (HSD) post-hoc test with the commonly used family-wise error rate (FWER) of $\alpha = 0.05$ to identify for which inconsistency level the alternative hypothesis, $H_{a,2}$, is supported. We hypothesize that if the quality of our manual labels is low, even small amounts of simulated labeling inconsistency could significantly reduce the model's performance, indicating support for $H_{a,2}$ across all levels.

\subsubsection{RQ3 Analysis Procedure} We compare pairwise distribution shifts between ReqSeek's predictions before (T0) and after transformation (T1--T6) instead of relying on performance metrics (precision, recall, $F_1$) as they aggregate per-class correctness, not relationship, before and after changes to the inputs. Therefore, we individually pair each transformation against T0 (T0$\times$T1, T0$\times$T2, \ldots, T0$\times$T6).
We employ the McNemar-Bowker test to evaluate pairwise distribution shifts in the model's predictions. McNemar-Bowker is a non-parametric test suitable for determining differences in distributions of paired or matched datasets for categorical data with more than two categories \cite{krampeBowkersTestSymmetry2007}. Further, to systematically test McNemar-Bowker, we define the following hypothesis, measure its magnitude with Cohen's $w$ effect size, and account for multiple tests using the Holm-Bonferroni method \cite{holmSimpleSequentiallyRejective1979}:

\begin{textbox}{\textit{ Hypothesis.}} \justifying
\noindent $H_{0,3}$: The distribution of predicted labels does not change before (T$_0$) and after (T$_i$) the transformation.
 
\noindent $H_{a,3}$: The transformation T$_i$ produces a significant change in the distribution.
\end{textbox}

Furthermore, we expect no support for $H_{a,3}$ for minor linguistic errors, while for major linguistic errors, we anticipate finding support for $H_{a,3}$ to achieve the desired robustness.

\subsubsection{RQ4 Analysis Procedure} 
We parsed each SRS at the sentence level and derived gold labels from the original document structure, assigning gold requirement labels to sentence-level units linked to requirement identifiers and auxiliary labels to all remaining units.

ReqSeek remains a three-class classifier throughout training and inference, predicting requirements, system-related auxiliaries, and contextual auxiliaries. However, the external SRSs provide independently derivable structures for identifying requirements, such as requirement identifiers or dedicated requirement sections, but not for distinguishing system-related from contextual auxiliary content. Therefore, the scope of the RQ4 OoD evaluation is constrained to requirement-versus-auxiliary identification under the labels available from the original SRS structures. Introducing auxiliary-subcategory labels ourselves would require subjective interpretation and would reduce the independence and fidelity of the OoD evaluation.

Consequently, we combine ReqSeek's two auxiliary predictions into a single auxiliary class only for RQ4 metric computation. This mapping changes only the evaluation granularity, not the model, training data, or training objective. The resulting binary scores are therefore not directly comparable to earlier three-class evaluations. The fine-grained three-class formulation remains important for the broader dataset-construction objective because it provides more informative labels for downstream AI tasks.

Since RQ4 assesses requirement identification across complete SRS documents, we report TP, FP, and FN counts alongside requirement-class performance metrics.  We additionally report prevalence and output fraction (Equations~\ref{eq:prevalence} and~\ref{eq:output_fraction}) to characterize requirement density within each SRS and the proportion of sentence-level units identified as requirements by ReqSeek.

Finally, the first author qualitatively characterized all FP and FN cases using the ARID annotation guide. Each disagreement was categorized as an omitted requirement, a false requirement prediction, a missed requirement, or a preprocessing issue. Omitted requirements are FPs that express requirements but were not listed as requirements in the original SRS-derived labels; false requirement predictions are FPs inspected as auxiliary; missed requirements are FNs inspected as requirements; and preprocessing issues are disagreements attributed to reference labels or sentence parsing rather than clear model errors. For characterization-adjusted precision, recall, and $F_1$ scores are computed, omitted requirements are counted as requirements, false requirement predictions remain FPs, and missed requirements remain FNs; preprocessing issues are excluded from model-miss counts. This characterization is used only to interpret disagreement sources and report secondary adjusted scores; the primary RQ4 metrics remain computed using the original SRS-derived labels.

\subsubsection{RQ5 Analysis Procedure} We analyze the results in two steps. First, to distinguish candidate items from actual requirements, we report corpus-level counts for SwaRD after applying ReqSeek. Specifically, we report the number of candidate items classified as requirements, system-related auxiliaries, and contextual auxiliaries.

Second, we characterize the SwaRD requirement subset with UOCSSR-ISO and the two human-curated reference datasets using aggregate textual and semantic metrics. The comparison focuses only on requirement instances because the human-curated datasets, namely PROMISE and NLP4RE, contain only requirements.

A direct instance‑by‑instance comparison between UOCSSR‑ISO and SwaRD is not feasible for two reasons: (i) SwaRD is constructed from all publicly available UOCSSR records, regardless of whether they were previously labeled as ISO 29148‑style requirements by the heuristic method; and (ii) the preprocessing for SwaRD changes the unit of analysis from heuristic‑derived records to sentence‑level candidate items. Therefore, we compare the resulting requirement subsets at the corpus level.

To measure the textual and semantic properties of the datasets, we use seven metrics: (1) dataset size, (2) average words per requirement, (3) average sentences per requirement, (4) vocabulary size, (5) lexical diversity, (6) semantic domain diversity, and (7) number of discovered topics.

Metrics 1--3 are straightforward counts normalized by dataset size. To address Metric 4, we lowercase words, remove numbers, discard duplicates, and count the remaining unique words.

Since the datasets' lengths vary, vocabulary size alone may understate diversity in smaller corpora. Therefore, we use Hypergeometric Distribution Diversity (HD-D) as metric 5. Intuitively, HD-D asks how likely a random 42-word draw is to contain varied words. A corpus repeatedly using the same words, such as ``system'', ``shall'', and ``should'', receives a lower score than one in which such a draw is more likely to include different words. Unlike type--token ratio, HD-D is less affected by text length because it estimates diversity from fixed-size random draws. HD-D is computed as:

\begin{align}
HD\text{-}D = \sum_{i=1}^{V} P(X_i \geq 1)
\end{align}

\noindent in which $V$ is the number of unique words and $P(X_i \geq 1)$ is the hypergeometric probability of observing at least one occurrence of word $i$ in a random draw of size $n$. Following \citet{mccarthyMTLDVocdDHDD2010}, we use a draw size of~42.

Finally, we operationalize metrics 6 and 7 using three BERTopic-based indicators \cite{220305794BERTopicNeural}. BERTopic is a transformer-based topic modeling approach that discovers topics from semantic text embeddings. Metric 6, domain diversity, is computed as the average normalized entropy over BERTopic-derived requirement-level topic distributions using:

\begin{align}
H(d_i)=-\sum_j p(t_j|d_i)\log p(t_j|d_i)
\end{align}

\noindent in which $d_i$ denotes requirement $i$, $t_j$ denotes topic $j$, and $p(t_j|d_i)$ is the probability that requirement $d_i$ is associated with topic $t_j$. Normalized entropy measures the evenness of each requirement's topic distribution, ranging from 0 to 1. Values closer to 1 indicate that requirements are associated with a more balanced mixture of semantic topics, while values near 0 indicate concentration around a few dominant topics. Metric 7, the number of discovered topics, is automatically determined by BERTopic. 

\subsection{Sample Size and Annotators}\label{ssec:sample_size}
We performed an a-priori power analysis to determine the sample sizes required for high statistical power for one-way ANOVA, McNemar-Bowker, and Friedman tests. Following \citet{dybaSystematicReviewStatistical2006}, we based the McNemar-Bowker and Friedman power analyses on their parametric analogues, one-way ANOVA and repeated-measures ANOVA, respectively. Using $\alpha = 0.05$, power $=0.8$, and an effect size of Cohen's $f = 0.5$ recommended for software engineering studies \cite{dybaSystematicReviewStatistical2006}, we calculated a required sample size of 53 observations in total across five groups for the one-way ANOVA. Using the same parameters, the ANOVA-based approximation for the McNemar–Bowker test yielded a required sample size of 42 observations. For the Friedman test, using the same $\alpha$ and power, along with Cohen's ANOVA effect size $f = 0.25$ recommended for software engineering studies \cite{dybaSystematicReviewStatistical2006}, the repeated-measures ANOVA approximation yielded a required sample size of 24 repeated-measures units.

For the exploratory human‑model evaluation, we estimated annotation sample size using a finite‑population correction on ARID's combined training and evaluation sets ($N = 2{,}396$; 1,916 training and 480 evaluation instances). With a 95\% confidence level, a 10\% margin of error, and a conservative $p = 0.5$, we obtained at least 93 instances. We rounded this value upward to 102 instances and drew a stratified sample of 34 instances per class from ARID's fixed evaluation set.

Three annotators independently labeled the 102 sampled instances. All annotators held an MSc degree and had professional backgrounds in software engineering or software development. Each annotator reported prior text or requirements annotation experience. Two annotators reported 3--5 years of software engineering experience and 1--3 years of requirements engineering experience, and one reported more than five years of software engineering experience and 3--5 years of requirements engineering experience. Two annotators reported moderate familiarity with SRS reading, specification, or writing, while one reported high familiarity. All three reported moderate familiarity with ISO 29148 or structured requirements.

\section{Results and Discussion}\label{sec:resanddisc}
We first present the results for each research question, followed by a brief discussion, and then summarize the findings for each objective.

\subsection{RQ1 Model Selection and Human Benchmarking}\label{ssec:rq1}
For the task-weighted $F_\beta$, we empirically estimated $\beta = 0.89$. We therefore report this metric as $F_{.89}$ to distinguish it from $F_1$.

\subsubsection{Candidate Model Selection Dimension}
Table~\ref{tab:model_comparison} reports the performance of the candidate models using 10-fold cross-validation. Except for GPT-2, all encoder-based models achieve comparable mean precision, recall, $F_1$, and $F_{.89}$ scores, indicating a similar average performance across 10 folds. GPT-2 performs notably worse, likely due to its generative architecture and pre-training objective, which may be less suited to the identification of discriminative requirements.

\begin{table}[ht]

\centering

\caption{10-Fold Cross-Validation Average Performance of the Candidate Models}
\label{tab:model_comparison}
\begin{tabular}{lcccc}
\hline
\textbf{Model} & \textbf{Precision} & \textbf{Recall} & \textbf{$F_{.89}$} & \textbf{$F_1$} \\
\hline
BERT & 0.94 $\pm$ 0.008 & 0.93 $\pm$ 0.011 & 0.94 $\pm$ 0.009 & 0.93 $\pm$ 0.010 \\
GPT-2 & 0.92 $\pm$ 0.013 & 0.90 $\pm$ 0.015 & 0.91 $\pm$ 0.014 & 0.91 $\pm$ 0.014 \\
all-MiniLM-L6-v2 & 0.94 $\pm$ 0.011 & 0.93 $\pm$ 0.018 & 0.94 $\pm$ 0.015 & 0.93 $\pm$ 0.015 \\
RoBERTa & 0.94 $\pm$ 0.008 & 0.93 $\pm$ 0.010 & 0.94 $\pm$ 0.008 & 0.93 $\pm$ 0.008 \\
\hline
\end{tabular}

\end{table}

The Friedman test applied to the fold-level precision scores obtained from the 10-fold cross-validation revealed a statistically significant difference among the evaluated models with a  $\chi^2(3) = 18.00$ and  $p < 0.001$, with a large effect size Kendall's $W = 0.60$ and 95\% confidence interval ranging from $[0.60, 0.74]$, supporting $H_{a,1}$.

Table \ref{tab:nemenyi_rq1} summarizes Nemenyi's post-hoc pairwise comparison test. The results show that GPT-2 performed significantly worse than BERT, all-MiniLM-L6-v2, and RoBERTa with an observed $p < 0.01$. However, no statistically significant differences were observed among BERT, all-MiniLM-L6-v2, and RoBERTa.

\begin{table}[h]
\centering
\caption{Nemenyi's Post-hoc Pairwise Comparison Results. Abbreviations: MRD, Mean Rank Difference}
\label{tab:nemenyi_rq1}
\begin{tabular}{llccc}
\hline
\textbf{Model 1} & \textbf{Model 2} & \textbf{MRD} & \textbf{p-adj} & \textbf{Reject} \\
\hline
GPT-2 & all-MiniLM-L6-v2 & 2.0 & 0.003 & True \\
GPT-2 & RoBERTa & 2.0 & 0.003 & True \\
GPT-2 & BERT & 2.0 & 0.003 & True \\
all-MiniLM-L6-v2 & RoBERTa & 0.0 & 1.000 & False \\
all-MiniLM-L6-v2 & BERT & 0.0 & 1.000 & False \\
RoBERTa & BERT & 0.0 & 1.000 & False \\
\hline
\end{tabular}
\end{table}

Therefore, BERT, all-MiniLM-L6-v2, and RoBERTa formed a statistically equivalent best-performing model group for requirements identification based on our evaluation data. From this group, we selected RoBERTa as the base model to fine-tune the final ReqSeek because of its optimized pre-training strategy and Byte Pair Encoding (BPE) tokenization, which might better handle out-of-vocabulary words, symbols, and technical tokens frequently occurring in SRS documents (e.g., identifiers, configuration parameters, and pseudo-code fragments). Table~\ref{tab:reqseek_per_class} reports ReqSeek's per-class performance scores on the fixed evaluation set, providing a class-level view of its performance.

\begin{table}[h]
\centering
\caption{ReqSeek Per-Class and Average Performance on the Fixed Evaluation Set}
\label{tab:reqseek_per_class}
\begin{tabular}{lccccc}
\hline
\textbf{Class} & \textbf{Precision} & \textbf{Recall} & \textbf{$F_{.89}$} & \textbf{$F_1$} & \textbf{Support} \\
\hline
Contextual auxiliary & 0.97 & 0.97 & 0.97 & 0.97 & 120 \\
Requirement & 0.95 & 0.96 & 0.95 & 0.96 & 210 \\
System-related auxiliary & 0.94 & 0.93 & 0.93 & 0.93 & 150 \\
\hline
Average & 0.95 & 0.95 & 0.95 & 0.95 & 480 \\
\hline
\end{tabular}
\end{table}

Further, Table~\ref{tab:empirical-comparison} contextualizes ReqSeek against representative ARI studies. Prior work reports best $F_1$ scores of 87\% \cite{bashirRequirementNotThat2023}, and 83\% \cite{wangAutomatedRequirementsDemarcation}, whereas our best-performing model group achieves an $F_1$ score of 95\%. Notably, even GPT-2 exceeds the reported scores of these prior studies. While these results are not directly comparable due to differences in datasets, domains, and evaluation protocols, they indicate that ARID is a promising basis for training requirements-identification models and motivate further evaluation under controlled cross-dataset settings.

\begin{table}[h]
\centering
\caption{Empirical Comparison with Representative Automatic Requirements Identification Studies Discussed in Section~\ref{sec:relatedw}. Abbreviations: P = Precision; R = Recall; CNN = Convolutional Neural Network; CV = Cross-Validation; TT = Train/Test Split; RFP = Request for Proposal;  OoD = Out-of-Distribution; LER = Linguistic-Error Robustness Evaluation; LIA = Labeling-Inconsistency Analysis; N/R = Not Reported}
\label{tab:empirical-comparison}
\resizebox{\textwidth}{!}{%
\begin{tblr}{
  width = \textwidth,
  colspec = {X[3.6cm,l] X[1.9cm,l] X[2.9cm,l] X[3.8cm,l] X[0.30cm,c] X[0.30cm,c] X[0.30cm,c]},
  row{1} = {font=\bfseries},
  hline{1,Z} = \heavyrulewidth,
  hline{2} = \lightrulewidth,
}
Study & Model & Domain/Artifact/Size & Evaluation Setting & P & R & $F_1$ \\ 
\citet{winklerAutomaticClassificationRequirements2016a} & CNN & Automotive; 90 specs\textsuperscript{\dag} & 90/10 split + 1 unseen spec  & 0.73 & 0.89 & N/R \\
\citet{falknerIdentifyingRequirementsRequests2019} & Naïve Bayes & Railway; 10 RFPs\textsuperscript{\dag} & 10‑fold CV + 4 held-out RFPs & N/R & 0.85 & N/R \\
\citet{abualhaijaMachineLearningBasedApproach2019} & Random Forest & Industrial; 33 specs\textsuperscript{\dag} & 10‑fold CV + unseen-specs eval. & 0.81 & 0.95 & N/R \\
\citet{ivanovExtractingSoftwareRequirements2022a} & BERT & PURE 30 SRSs & Train/val/test & 0.92 & 0.80 & 0.86 \\
\citet{bashirRequirementNotThat2023} & BERT & Railway; 5 tenders docs\textsuperscript{\dag} & 5‑fold CV & N/R & N/R & 0.87 \\
\citet{wangAutomatedRequirementsDemarcation} & DeBERTa & PURE 30 SRSs + Dronology & 5‑fold CV + 70/30 TT & N/R & N/R & 0.83 \\
\citet{vancanLocatingRequirementsBacklog2025} & RoBERTa & Agile; 14 projects\textsuperscript{\dag} & 7-fold project-level CV & N/R & N/R & 0.82 \\
\textbf{ReqSeek (this study)} & RoBERTa & ARID; 23 SRSs from different domains & 10‑fold CV + 80/20 TT + OoD (3 SRS) + LER + LIA & 0.95 & 0.95 & 0.95 \\
\end{tblr}
}
\begin{flushleft}
\textsuperscript{\dag}Artifacts/data were fully or partly used only within the original study and not publicly released.
\end{flushleft}
\end{table}

\subsubsection{Exploratory Human-Model Benchmarking}
Of the 102 sampled instances, one instance produced complete disagreement among the three independent annotators. Because no majority-vote consensus label could be derived for this instance, it was excluded from the consensus-based agreement and benchmarking analyses. The remaining 101 instances were used for the exploratory human-model benchmarking.

Table~\ref{tab:rq1_agreement} summarizes the agreement and alignment results for ReqSeek, the first-author labels, and the three independent annotators. Fleiss' $\kappa$ among the three independent annotators was 0.67, indicating substantial agreement. This suggests that the ARI task can be performed consistently by independent annotators, while still involving a non-trivial degree of subjective judgment.

The comparison of each labeling source against the consensus labels showed consistently high agreement. Cohen's $\kappa$ ranged from 0.80 to 0.86 across all sources, corresponding to substantial to almost perfect agreement with the consensus reference. The highest agreement with the consensus was achieved by Annotator A2 with $\kappa=0.86$, followed by Annotator A1 with $\kappa=0.85$, the first-author labels with $\kappa=0.84$, ReqSeek with $\kappa=0.81$, and Annotator A3 with $\kappa=0.80$. Overall, ReqSeek's agreement with the consensus falls within the range observed among independent human annotators, indicating that its labeling behavior aligns closely with human judgments. The strong agreement of the first-author ARID labels with the consensus further suggests that the ARID labeling task is reasonably consistent with independent human judgments.

\begin{table}[h]
\centering
\caption{Agreement and Alignment Results for the Exploratory Human-Model Benchmarking}
\label{tab:rq1_agreement}
\begin{tabular}{lccc}
\hline
\textbf{Source} & \textbf{Compared Instances} & \textbf{Cohen's $\kappa$ vs. Consensus} & \textbf{Fleiss' $\kappa$} \\
\hline
First-author Labels & 101 & 0.84 & -- \\
ReqSeek Predictions & 101 & 0.81 & -- \\
Annotator A1 & 101 & 0.85 & -- \\
Annotator A2 & 101 & 0.86 & -- \\
Annotator A3 & 101 & 0.80 & -- \\
\hline
Annotators A1--A3 & 101 & -- & 0.67 \\
\hline
\end{tabular}
\end{table}

Table~\ref{tab:rq1_human_model_benchmark} reports the benchmarking results against the consensus labels using standard classification metrics. Overall, all comparison sources achieve similar performance levels across precision, recall, $F_1$, and $F_{.89}$. Precision ranges from 0.89 to 0.92, with the first-author ARID labels achieving the highest precision at 0.92, followed by Annotator A2 at 0.91, ReqSeek and Annotator A1 at 0.90, and Annotator A3 at 0.89. ReqSeek achieves precision within the range of the independent annotators, with an output fraction close to the consensus prevalence, indicating that its predictions are broadly aligned with human labeling behavior on the ternary ARID labeling task.

\begin{table}[h]
\centering
\caption{Exploratory Human-Model Benchmarking Against Consensus Labels on the Ternary ARID Labeling Task. Abbreviations: P = precision; R = recall; PRV = Prevalence; OF = Output Fraction; TP Time = Mean Time to Identify True Positives; FP Time = Mean Time to Decide False Positives}
\label{tab:rq1_human_model_benchmark}
\resizebox{\textwidth}{!}{%
\begin{tabular}{lccccccccc}
\hline
\textbf{Source} &\textbf{P} & \textbf{R} & \textbf{$F_1$} & \textbf{$F_{.89}$} & \textbf{PRV} & \textbf{OF} & \textbf{TP Time} & \textbf{FP Time} & \textbf{Total Time} \\
\hline
First-author Labels & 0.92 & 0.89 & 0.90 & 0.90 & 0.37 & 0.34 & -- & -- & -- \\
ReqSeek Predictions & 0.90 & 0.87 & 0.88 & 0.88 & 0.37 & 0.32 & 0.01 (sec) & 0.01 (sec) & 0.02 (min) \\
Annotator A1 & 0.90 & 0.90 & 0.90 & 0.90 & 0.37 & 0.37 & 17.2 (sec) & 33.9 (sec) & 31.7 (min)\\
Annotator A2 & 0.91 & 0.91 & 0.91 & 0.91 & 0.37 & 0.35 & 18.2 (sec) & 20.3 (sec) & 31.0 (min)\\
Annotator A3 & 0.89 & 0.87 & 0.87 & 0.88 & 0.37 & 0.30 & 30.4 (sec) & 23.9 (sec) & 49.7 (min)\\
\hline
\end{tabular}
}
\end{table}

The timing results reveal a substantial efficiency difference between human annotation and automated prediction using ReqSeek. The three independent annotators required between 31.0 and 49.7 minutes to label the full sampled instances. Their mean times to find true positives ranged from 17.2 to 30.4 seconds, while their mean times to decide false positives ranged from 20.3 to 33.9 seconds. Based on these timings, the ratios for Annotators A1, A2, and A3 were 0.507, 0.898, and 1.269, respectively. Their average yields $\beta = (0.507 + 0.898 + 1.269) / 3 \approx 0.89$, as reported at the beginning of Section \ref{ssec:rq1}. In contrast, ReqSeek processed the same sample in 0.02 minutes, with mean times of 0.01 seconds for both finding true positives and deciding false positives. Although ReqSeek's timing is hardware-specific and therefore not directly generalizable, these results indicate that automated ARI can substantially reduce prediction time compared to manual annotation while maintaining performance comparable to that of human annotators. However, the time required for humans to review ReqSeek's output and reject false positives was not measured.

Together, these results show that ReqSeek falls within the range of independent human annotators in agreement and ternary classification performance on the sampled evaluation data, while requiring substantially less prediction time. Accordingly, the benchmark provides a human reference point for interpreting ReqSeek's in-distribution performance.

\noindent \textbf{\textit{By identifying BERT, all-MiniLM-L6-v2, and RoBERTa as a statistically equivalent top-performing group, selecting RoBERTa as the foundation model for ReqSeek, demonstrating human-comparable labeling agreement and classification performance together with substantially lower prediction time, and providing evidence of ARID labeling consistency, we fulfill \mbox{Objective 1}.}}

\subsection{RQ2 Labeling Inconsistency Analysis}
The descriptive statistics for RoBERTa-base, the selected baseline model for ReqSeek, are computed across five inconsistency levels, each repeated five times. Mean precision remains relatively stable from 0\% to 20\% inconsistency, decreasing from $\hat{x}=0.95$ at 0\% to $\hat{x}=0.94$ at 10\% and $\hat{x}=0.92$ at 20\%, with $s=0.008$ at all three levels. Beyond 20\%, performance declines more noticeably, reaching $\hat{x}=0.88$ at 30\% and $\hat{x}=0.86$ at 40\%, while variability increases to $s=0.017$ and $s=0.026$, respectively.

Table~\ref{tab:rq2_anova_assumptions} summarizes the assumption checks for the one-way ANOVA. Shapiro-Wilk tests did not indicate significant deviations from normality across inconsistency levels (all $p > .05$; $W = 0.8887$--$0.9947$). Levene's test also did not indicate heterogeneity of variances ($F = 2.0082$, $p = .1321$). Therefore, one-way ANOVA was considered appropriate.

\begin{table}[h]
\centering
\caption{One-way ANOVA Assumption Checks for RQ2.}
\label{tab:rq2_anova_assumptions}
\begin{tabular}{lcc}
\toprule
\textbf{Test} & \textbf{Statistic} & \textbf{$p$-value} \\
\midrule
Shapiro--Wilk (0\% Noise)  & $W = 0.9947$ & 0.9934 \\
Shapiro--Wilk (10\% Noise) & $W = 0.9690$ & 0.8685 \\
Shapiro--Wilk (20\% Noise) & $W = 0.8887$ & 0.3506 \\
Shapiro--Wilk (30\% Noise) & $W = 0.9380$ & 0.6520 \\
Shapiro--Wilk (40\% Noise) & $W = 0.9613$ & 0.8168 \\
Levene's Test              & $F = 2.0082$ & 0.1321 \\
\bottomrule
\end{tabular}
\end{table}
Further, one-way ANOVA results yielded an $F(4, 20) = 25.30$ and $p < 0.01$ with a large effect size $\eta^2_p = 0.8350$, indicating that at least one level of inconsistency significantly impacted model performance, supporting $H_{a,2}$. The 95\% confidence interval for $\eta^2_p$ ranged from $[0.77, 0.94]$.

Tukey's HSD (see Table~\ref{tab:rq2_hsd}) shows a significant precision drop at $30\%$ inconsistency, with $0\%$ vs.~$30\%$ and $40\%$ having $p < 0.01$. In contrast, comparisons between $0\%$ and $10\%$ or $20\%$ yield $p > 0.05$, indicating no significant precision drop below $20\%$ inconsistency. Thus, we fail to reject the null hypothesis for these levels.

\begin{table}[ht]
\centering
\caption{Multiple Comparisons of Precision Means Using Tukey's HSD Test (FWER~=~0.05)}
\label{tab:rq2_hsd}
\begin{tblr}{
  columns = {c},
  hline{1-2,12} = {-}{},
}
\textbf{Group 1} & \textbf{Group 2} & \textbf{Mean Diff} & \textbf{p-adj} & \textbf{Lower} & \textbf{Upper} & \textbf{Reject}\\
0\% & 10\% & -0.0060 & 0.9722 & -0.0356 & 0.0236 & False\\
0\% & 20\% & -0.0244 & 0.1384 & -0.0539 & 0.0052 & False\\
0\% & 30\% & -0.0603 & 0.0001 & -0.0899 & -0.0308 & True\\
0\% & 40\% & -0.0810 & 0.0001 & -0.1106 & -0.0515 & True\\
10\% & 20\% & -0.0184 & 0.3696 & -0.0479 & 0.0112 & False\\
10\% & 30\% & -0.0543 & 0.0002 & -0.0839 & -0.0248 & True\\
10\% & 40\% & -0.0750 & 0.0001 & -0.1046 & -0.0455 & True\\
20\% & 30\% & -0.0360 & 0.0125 & -0.0655 & -0.0064 & True\\
20\% & 40\% & -0.0567 & 0.0001 & -0.0862 & -0.0271 & True\\
30\% & 40\% & -0.0207 & 0.2603 & -0.0503 & 0.0089 & False
\end{tblr}
\end{table}

These results imply that poorer initial labels would degrade performance even at $10\%$ or $20\%$ inconsistency -- levels substantial enough to harm model training. These findings indicate that the negligible performance drop below $20\%$ underscores both the selected model's resilience to imperfect labels and the consistency of our manual labeling.

\subsection{RQ3 Linguistic Errors Robustness Evaluation}
Table~\ref{tab:robustness_results} presents ReqSeek's performance under T0 (without transformation) and after each transformation (T1--T6), and Table~\ref{tab:stat_results} summarizes McNemar statistics. 

\begin{table}[h]
\centering
\caption{\textnormal{ReqSeek} Performance Under Linguistic Errors}
\label{tab:robustness_results}
\begin{tblr}{
  colspec = {lccccccc},
  hline{1,6} = {-}{0.08em},
  hline{2} = {-}{},
}
 & \textbf{T0} & \textbf{T1} & \textbf{T2} & \textbf{T3} & \textbf{T4} & \textbf{T5} & \textbf{T6}\\
\textbf{Severity} & - & Minor & Minor & Minor & Major & Major & Major\\
Precision & 0.95 & 0.94 & 0.92 & 0.92 & 0.79 & 0.90 & 0.78\\
Recall & 0.95 & 0.94 & 0.92 & 0.93 & 0.65 & 0.90 & 0.62\\
\textbf{$F_1$} & 0.95 & 0.94 & 0.92 & 0.93 & 0.53 & 0.90 & 0.50
\end{tblr}
\end{table}

\begin{table}[h]
\centering
\caption{\textnormal{ReqSeek} Comparison on Linguistic Errors. Abbreviations: $p$ = $p$-value; Adj. $p$ = Holm-Bonferroni Adjusted $p$-value; $w$ = Cohen's $w$ Effect Size; $CI$ = Confidence Interval}
\label{tab:stat_results}
\begin{tblr}{
  width = \linewidth,
  colspec = {Q[360]Q[63]Q[65]Q[65]Q[50]Q[98]Q[73]Q[115]},
  row{1} = {c},
  cell{2}{3} = {c},
  cell{2}{4} = {c},
  cell{2}{5} = {c},
  cell{2}{6} = {c},
  cell{2}{7} = {c},
  cell{2}{8} = {c},
  cell{3}{3} = {c},
  cell{3}{4} = {c},
  cell{3}{5} = {c},
  cell{3}{6} = {c},
  cell{3}{7} = {c},
  cell{3}{8} = {c},
  cell{4}{3} = {c},
  cell{4}{4} = {c},
  cell{4}{5} = {c},
  cell{4}{6} = {c},
  cell{4}{7} = {c},
  cell{4}{8} = {c},
  cell{5}{3} = {c},
  cell{5}{4} = {c},
  cell{5}{5} = {c},
  cell{5}{6} = {c},
  cell{5}{7} = {c},
  cell{5}{8} = {c},
  cell{6}{3} = {c},
  cell{6}{4} = {c},
  cell{6}{5} = {c},
  cell{6}{6} = {c},
  cell{6}{7} = {c},
  cell{6}{8} = {c},
  cell{7}{3} = {c},
  cell{7}{4} = {c},
  cell{7}{5} = {c},
  cell{7}{6} = {c},
  cell{7}{7} = {c},
  cell{7}{8} = {c},
  cell{8}{1} = {c=8}{0.929\linewidth},
  hline{1,9} = {-}{0.08em},
  hline{2,8} = {1,3-8}{0.03em},
  hline{2,8} = {2}{},
}
\textbf{Comparison} & \textbf{Severity} & $p$ & \textbf{Adj.~$p$} & $w$ & \textbf{95\% $CI$} & \textbf{Significant} & $\chi^2_{Bowker}$\\
T0$\times$T1: Add Minor Details & Minor & $0.096$ & $0.261$ & 0.115 & {[}0.108,~ 0.122] & No & 6.3444\\
T0$\times$T2:~Re-order Words & Minor & $0.087$ & $0.261$ & 0.117 & {[}0.110,~ 0.124] & No & 6.5647\\
T0$\times$T3:~Remove Minor Details~ & Minor & $0.393$ & $0.393$ & 0.079 & {[}0.074,~ 0.084] & No & 2.9897\\
{T0$\times$T4: Signaling Keywords Removed} & Major & $<.001$ & $<.001$ & 0.658 & {[}0.616,~ 0.699] & Yes & 207.6714\\
{T0$\times$T5: ISO Syntax Violation\\ (+Signaling Keywords)} & Major & $<.001$ & $0.002$ & 0.189 & {[}0.177,~ 0.201] & Yes & 17.2177\\
{T0$\times$T6: ISO Syntax Violation\\(-Signaling Keywords)} & Major & $<.001$ & $<.001$ & 0.666 & {[}0.624,~ 0.708] & Yes & 212.7450\\
\textbf{Sample size} ($n$): T0 = 480, Ti = 480; predicted class distribution for T0: requirement = 213, contextual auxiliary = 119, system-related auxiliary = 148. &  &  &  &  &  &  & 
\end{tblr}
\end{table}

For three minor linguistic error transformations (T1--T3), the McNemar test yielded $p > 0.05$, failing to reject $H_{0,3}$. This suggests ReqSeek's predictions remain stable under minor textual variations, as was our aim.
In contrast, for the major linguistic error transformations (T4--T6), the McNemar test yielded a $p < 0.01$ with a large effect size $w = 0.65$ for T4 and $ w = 0.66$ for T6, and a small effect size $w = 0.18$ for T5. These results strongly support $H_{a,3}$, demonstrating that major linguistic errors significantly alter ReqSeek's predictions aligned with our objective.

These results confirm ReqSeek's intended behavior: insensitive to minor linguistic errors but sensitive to major errors, effectively filtering high-quality requirements for standardized datasets.

\subsection{RQ4 Domain-Agnosticism Evaluation}
Table~\ref{tab:rq4_ood_results} presents ReqSeek's requirement-identification performance on three previously OoD SRSs. ReqSeek achieves requirement-class precisions of 88\%, 59\%, and 94\% for NTP, RMOS, and KIB\textsuperscript{3}, respectively. The high requirement-class recall further indicates strong requirements coverage across all three SRSs. However, the precision of the predicted requirement sets varies across sources, particularly for RMOS. 

\begin{table}[h]
\centering
\caption{ReqSeek OoD Requirement-Identification Results on Three Authentic SRSs. Abbreviations: $P_\mathrm{req}$ = Requirement-Class Precision; $R_\mathrm{req}$ = Requirement-Class Recall; $F_{1,\mathrm{req}}$ = Requirement-Class $F_1$; Prev. = Prevalence of Gold-Labeled Requirements; OF = Fraction of Input Sentences Predicted as Requirement}
\label{tab:rq4_ood_results}
\begin{tabular}{lccccccccccc}
\toprule
\textbf{SRS} & \textbf{Input $N$} & \textbf{Gold req.} & \textbf{Pred. req.} & \textbf{TP} & \textbf{FP} & \textbf{FN} & $\mathbf{P_\mathrm{req}}$ & $\mathbf{R_\mathrm{req}}$ & $\mathbf{F_{1,\mathrm{req}}}$ & \textbf{Prev.} & \textbf{OF} \\
\midrule
NTP  & 1065 & 172 & 194 & 171 & 23 & 1 & 0.88 & 0.99 & 0.93 & 0.16 & 0.18 \\
RMOS & 724  & 70  & 117 & 69  & 48 & 1 & 0.59 & 0.99 & 0.74 & 0.10 & 0.16 \\
KIB\textsuperscript{3} & 296  & 35  & 34  & 32  & 2  & 3 & 0.94 & 0.91 & 0.93 & 0.12 & 0.11 \\
\bottomrule
\end{tabular}
\end{table}

For NTP and RMOS, ReqSeek's output fraction exceeds the prevalence of gold-labeled requirements: 18\% versus 16\% for NTP and 16\% versus 10\% for RMOS. Combined with high recall in both cases, this suggests that ReqSeek identifies additional requirement candidates not present in the original SRS-derived labels. The discrepancy is small for NTP, where precision remains at 88\%, but is more noticeable for RMOS, where the gap between prevalence and output fraction is larger, and precision is 59\%. In contrast, KIB\textsuperscript{3} exhibits a slightly lower output fraction than its prevalence (11\% versus 12\%), indicating a more selective prediction set. This selectivity is reflected in slightly lower recall relative to precision, as three gold-labeled requirements are misclassified. Thus, KIB\textsuperscript{3} represents a cleaner but more selective case, while NTP and especially RMOS illustrate scenarios in which ReqSeek identifies additional requirement candidates beyond the original labels. Table~\ref{tab:rq4_error_characterization} summarizes the qualitative characterization of the false-positive and false-negative cases.

\begin{table}[h]
\centering
\caption{Qualitative Error Characterization of RQ4 FP and FN cases. Abbreviations: Omitted Req. = Omitted Requirements; False Req. Pred. = False Predicted Requirements; Missed Req. Missed Requirements; Prep. Issues = Preprocessing Issues;  $P_{adj}$ = Adjusted Precision, $R_{adj}$ = Adjusted Recall; $F1_{adj}$ = Adjusted $F_1$}
\label{tab:rq4_error_characterization}
\begin{tabular}{lccccccccc}
\toprule
\textbf{SRS} & \textbf{FP} & \textbf{Omitted Req.} & \textbf{False Req. Pred.} & \textbf{FN} & \textbf{Missed Req.} & \textbf{Prep. Issues} & \textbf{$P_{adj}$} & $R_{adj}$ &$F1_{adj}$\\
\midrule
NTP  & 23 & 13 & 10 & 1 & 0 & 1 & 0.95 & 1.00 & 0.97\\
RMOS & 48 & 32 & 16 & 1 & 0 & 1 & 0.86 & 1.00 & 0.93\\
KIB\textsuperscript{3} & 2  & 1  & 1  & 3 & 3 & 0 & 0.97 & 0.92 & 0.94\\
\bottomrule
\end{tabular}
\end{table}

Across a total of 2085 sentence-level units from the three OoD SRSs, ReqSeek misclassified 78 cases. Qualitative inspection revealed that, in NTP, 13 of 23 false positives were omitted requirements, while 10 were false requirement predictions. In RMOS, 32 of 48 false positives were omitted requirements, and 16 were false requirement predictions. Thus, many of the additional candidates identified by ReqSeek, particularly in RMOS, expressed requirements not listed in the original SRS-derived labels, rather than non-requirements incorrectly predicted as requirements.

Using this characterization-adjusted view, $P_\mathrm{req}$ increases from 0.88 to 0.95 for NTP, from 0.59 to 0.86 for RMOS, and from 0.94 to 0.97 for KIB\textsuperscript{3}. The corresponding adjusted $F_1$ values are 0.97, 0.93, and 0.94, respectively. These adjusted scores do not replace the original SRS-derived metrics in Table~\ref{tab:rq4_ood_results}; rather, they show that measured precision, especially for RMOS, is strongly affected by omitted requirements in the original SRS-derived labels.

Overall, the results for RQ4 indicate that ReqSeek can identify requirements in out-of-domain SRSs. The qualitative characterization further demonstrates that the majority of ReqSeek's predicted requirements are valid requirement statements upon inspection, including in the initially lowest-precision RMOS case. Notably, the omitted requirements were not limited to trivial formatting disagreements; they included statements related to functional behavior, access control, security, audit-related constraints, interface behavior, and user interface properties. Therefore, RQ4 suggests ReqSeek's suitability for constructing high-quality requirements datasets. The complete characterization file is included in the replication package, and representative examples are presented in Appendix~\ref{app:rq4_examples}.

\subsection{RQ5 Use-Case Validation}
Table~\ref{tab:sward_class_distribution} summarizes the class distribution of SwaRD after applying ReqSeek. ReqSeek identified 10,201 requirement instances out of 40,522 in SwaRD, making up 25.2\%.

\begin{table}[h]
\centering
\caption{Class Distribution of SwaRD After Applying ReqSeek}
\label{tab:sward_class_distribution}
\begin{tabular}{lrr}
\toprule
\textbf{Class} & \textbf{Count} & \textbf{Percentage} \\
\midrule
Requirement & 10,201 & 25.2\% \\
System-related Auxiliary & 25,746 & 63.5\% \\
Contextual Auxiliary & 4,575 & 11.3\% \\
\midrule
Total instances & 40,522 & 100\% \\
\bottomrule
\end{tabular}
\end{table}

Table~\ref{tab:req_stats} presents textual and semantic properties for the requirements subset of SwaRD, UOCSSR-ISO, and the two human-curated reference datasets.

\begin{table}[h]
\centering
\caption{Textual and Semantic Diversity Properties for Requirements Datasets}
\label{tab:req_stats}
\begin{tabular}{lrrrr}
\toprule
\textbf{Metric}  & \textbf{UOCSSR-ISO} & \textbf{SwaRD} & \textbf{PROMISE} & \textbf{NLP4RE} \\
\midrule
Dataset Size                    & 12,080 & 10,201 & 969     & 212 \\
Avg. Words/Requirement      & 83.93  & 20.19  & 19.31   & 32.68 \\
Avg. Sentences/Requirement  & 3.30   & 1.00   & 1.18    & 1.00 \\
Vocabulary Size                 & 11,170 & 8,148  & 2,010   & 1,120 \\
Lexical Diversity               & 0.87   & 0.83   & 0.79    & 0.80 \\
Semantic Domain Diversity       & 0.56   & 0.64   & 0.57    & 0.30 \\
Number of Discovered Topics     & 260    & 174    & 26      & 2 \\
\bottomrule
\end{tabular}
\end{table}

UOCSSR requirements are considerably longer than those in the human-curated PROMISE and NLP4RE datasets, averaging 83.93 words and 3.3 sentences per requirement. In contrast, SwaRD yields concise, atomic requirements, averaging 20.19 words and 1.0 sentences per requirement, similar to the textual properties of PROMISE and NLP4RE. 

The atomic sentence properties of SwaRD relative to UOCSSR-ISO are expected because SwaRD is constructed from sentence-level instances. Therefore, this difference should not be interpreted as an isolated effect of the model alone. Rather, it reflects the combined outcome of sentence-level preprocessing and ReqSeek classification. At the corpus level, the resulting requirement subset more closely resembles the ISO 29148-style requirement statements discussed in Section \ref{ss:isosrs}, where each requirement should express a single capability, specify a subject, assign responsibility to a system or component, define the expected or permitted action, and include relevant conditions or constraints within a single sentence.

UOCSSR‑ISO exhibits the largest vocabulary size of 11,170, a lexical diversity of 0.87, and the highest number of discovered topics at 260. Despite these higher values, its semantic domain diversity score of 0.56 is lower than those of SwaRD at 0.64 and PROMISE at 0.57. Because this metric reflects the evenness of topic distributions within individual requirements, the combination of a lower domain diversity score with longer records, greater vocabulary size, and more topics suggests that UOCSSR-ISO may capture more heterogeneous textual material, not only well-bounded requirement statements.

The lower semantic domain diversity values for PROMISE and NLP4RE relative to SwaRD should not be interpreted as evidence of verbosity or heterogeneous records, given that these datasets are smaller and contain fewer topics. These values must also be considered alongside their narrower scope, which includes vocabulary sizes of 2,010 and 1,120, lexical diversity scores of 0.79 and 0.80, and 26 and 2 discovered topics, respectively.

Although SwaRD is as concise as the PROMISE and NLP4RE datasets, it exceeds them with 10,201 requirements, a larger vocabulary of 8,148 unique words, and a lexical diversity of 0.83, offering a large-scale requirements dataset for AI applications in requirements engineering research.

\noindent \textit{\textbf{Objective 2 is fulfilled by: (1) rigorously validating our manual labeling quality, (2) confirming ReqSeek’s stability under minor linguistic variations, (3) evaluating ReqSeek on OoD SRSs to assess requirement identification beyond the training corpus and provide evidence against overfitting to ARID, and (4) validating its use for requirements engineering research by contributing SwaRD, a large-scale dataset whose requirement subset shows textual and semantic properties comparable to human-curated datasets while extending them in size and semantic coverage.}}

\section{Implications}\label{sec:implic}
This section examines the broader significance of our contributions: ARID, ReqSeek, and SwaRD.

\subsection{Research Implications}
ReqSeek addresses a significant gap in the largely manual identification of requirements from SRSs by fine-tuning transformer-based models on heterogeneous SRSs, enabling domain-agnostic ARI for standardized dataset creation to support requirements engineering research. ReqSeek's domain-agnostic requirement-identification capability is supported both by design through ARID's cross-domain composition and empirically through OoD evaluation on unseen SRSs with requirements independently labeled by their original authors.

Further, ARID not only enables ReqSeek’s requirements identification domain-agnosticism capability but also suggests that a well-constructed dataset can be sufficient to achieve robust performance. Only if such a dataset fails to generalize should efforts turn to complex pipelines or ad-hoc model tweaks, as explored in prior research.

We evaluate ReqSeek's practicality and contribute SwaRD, whose textual and semantic properties align with human-curated datasets, while scaling them in size and domain coverage. SwaRD fills a significant gap in research applying AI to requirements engineering by constructing a large-scale dataset with 10,201 requirement instances. Future work can extend \mbox{ReqSeek} beyond ISO 29148-style to additional requirement types, building on ARID and our method for continued collaboration.

\subsection{Practical Implications}
While we focus on ReqSeek’s use case in constructing high-quality requirements datasets for research, its domain-agnostic design and robustness suggest potential for practical industry applications. However, these applications were not directly evaluated in this study. Related work has applied models similar to ReqSeek to oversight reduction and risk assessment \cite{falknerIdentifyingRequirementsRequests2019, bashirRequirementNotThat2023, winklerAutomaticClassificationRequirements2016a}, traceability \cite{hayesWekaMeetsTraceLab2014, abualhaijaMachineLearningBasedApproach2019}, and hidden requirements discovery, as also illustrated by the RMOS cases in RQ4.

\section{Threats to Validity}\label{sec:limit}
This section discusses potential threats to the validity of our study and the measures taken to mitigate them.

\subsection{Construct Validity}
The ARID dataset was manually labeled by the first author. While involving multiple annotators across the full dataset would have enabled inter-rater reliability analysis and provided additional evidence for label quality, prior work suggests that a single coder operating under clearly defined class boundaries can improve labeling consistency and reduce interpretive variation~\cite{DifferentResearchersDifferent}. Consequently, our study reflects a tradeoff between full-dataset multi-annotator agreement and single-coder consistency.

To mitigate this threat, we conducted an exploratory human-model benchmark on a stratified sample drawn from the fixed 20\% ARID evaluation split, which remained unseen by ReqSeek during training. Three independent annotators labeled each sample, and consensus labels were determined by majority vote. We measured agreement among annotators using Fleiss’ $\kappa$ and agreement with the consensus using Cohen’s $\kappa$ for each annotator, the first author, and ReqSeek. 

To further evaluate the robustness of our measurement approach, we simulated increasing levels of label inconsistency and analyzed ReqSeek's sensitivity to human labeling errors. This analysis provides additional evidence regarding the stability of the classification task under imperfect annotations.

Another threat to construct validity concerns the interpretation of the RQ5 dataset comparison. Because UOCSSR and SwaRD were created using different construction methodologies, the results should not be interpreted as evidence of the superiority of either approach. Instead, the comparison is intended solely to characterize differences in the datasets’ textual and semantic properties.

\subsection{Internal Validity}
Differences in annotators' interpretations of the labeling criteria may have influenced the human-model benchmark. To reduce this threat, we recruited annotators with experience in requirements engineering, familiarity with SRS development and analysis, and knowledge of ISO 29148 requirements. Annotators received onboarding materials, labeling guidelines, and a knowledge refresher, and were allowed to familiarize themselves with the labeling tool before performing the task.

Additionally, RQ2 and RQ3 were evaluated using the same held-out test set. This design choice ensured that observed performance differences could be attributed to the investigated experimental conditions, such as model architecture, label inconsistencies, or textual variations, rather than differences in evaluation data. Consequently, comparability across experiments was strengthened.

\subsection{External Validity}
We evaluated ReqSeek on three unseen real-world SRS documents, whose labels were independently assigned by their original authors. However, these SRS structures only support the distinction between requirements and non-requirements. As a result, the OoD evaluation was performed at the binary level by merging ReqSeek's system-related and contextual auxiliary class predictions into a single general auxiliary class. Therefore, the results of RQ4 should be interpreted as evidence of domain-agnostic requirement identification rather than evidence of full three-class OoD performance.

Our categorization of minor and major linguistic errors, while grounded in practical considerations, may not represent the full range of linguistic issues encountered in industrial requirements specifications. Since we are the first to investigate the effects of such textual variations on requirements classification, established taxonomies and guidelines remain limited. Future work should explore a broader range of error types and alternative categorizations.

Finally, although the annotators involved in the human-model benchmark possessed relevant domain knowledge and experience, the study included only three participants. Consequently, the results may not fully reflect the diversity of requirements engineering practitioners, limiting the generalizability of the observed human performance and agreement levels.

\subsection{Conclusion Validity}
For all inferential statistical analyses, we conducted a priori power analyses for the Friedman, ANOVA, and McNemar--Bowker tests. Following \citet{dybaSystematicReviewStatistical2006}, the required sample sizes for the non-parametric tests were approximated using their parametric counterparts.

The resulting sample size was sufficient for the McNemar–Bowker test under the adopted approximation, whereas the Friedman analysis used ten folds compared to the estimated requirement of 24. However, the ANOVA analysis was limited to five observations per group. Achieving the recommended sample size of 11 observations per group would have required 55 model training sessions, which exceeded available computational resources. To support the use of ANOVA despite this limitation, we verified the assumptions of normality and homogeneity of variance using the Shapiro-Wilk and Levene's tests, respectively.

The exploratory human-model evaluation was conducted on a stratified sample, with the sample size determined using finite-population sampling at the 95\% confidence level. Its purpose was to provide an indicative benchmark of ReqSeek’s performance relative to human judgments and effort estimates. While the selected sample was sufficient for this exploratory analysis, a larger sample could provide more precise estimates of agreement and performance. Additionally, the limited number of annotators may constrain the strength and generalizability of the conclusions. Consequently, the benchmark should be interpreted as an exploratory comparison rather than a comprehensive human-subject study of requirements analysts. Future work involving a larger and more diverse participant pool, richer task settings, and qualitative analyses of human decision-making would strengthen these findings.

\section{Conclusion}\label{sec:conc}
We introduced ReqSeek, a fine-tuned, domain-agnostic, transformer-based model for automatic ISO 29148-style requirements identification from heterogeneous SRSs.

To support ReqSeek's domain-agnosticism, we constructed the Automatic Requirements Identification Dataset (ARID) with 2,396 manually reviewed and labeled instances from 23 heterogeneous SRSs. Inter-rater agreement analysis with independent annotators provided evidence of ARID’s label quality. After fine-tuning four open-source transformer models, RoBERTa-base was selected as the base model for ReqSeek and achieved 95\% precision, recall, $F_1$, and $F_{.89}$ on the held-out evaluation set. Exploratory human-model benchmarking further showed performance comparable to human annotators in requirements identification while requiring substantially less prediction time than manual annotation.

Evaluation demonstrates ReqSeek’s robustness and supports its applicability beyond ARID. With up to 20\% random training data mislabeling, precision scores remained stable, declining gradually beyond that threshold. Minor linguistic errors did not affect the prediction distribution or performance, whereas major ISO 29148-style violations produced expected drops, confirming the reliance on standard syntax.

To assess domain-agnostic requirement identification, we conducted a post-hoc OoD evaluation on unseen SRSs with independently labeled requirements. The results provide evidence that ReqSeek identifies requirements across unseen SRSs, while qualitative error characterization indicates that many apparent false positives corresponded to valid requirements omitted from the original SRS-derived labels, suggesting that measured precision may underestimate actual performance. These findings support reliable requirement identification beyond ARID and provide additional evidence for ARID's label quality.

ReqSeek also contributes SwaRD, containing 10,201 automatically identified requirements with textual and semantic properties aligned with human-curated datasets while providing substantially greater scale and domain diversity.

In summary, our results demonstrate the feasibility of robust, domain-agnostic transformer-based requirements identification across heterogeneous SRSs, enabling automatic construction of large-scale requirements datasets.

Future work will expand ReqSeek to additional requirement types (agile user stories, acceptance test cases) and grow ARID and SwaRD to facilitate broader AI applications and research.

\section{Data Availability and Licensing}\label{sec:da}

The replication package for this study, including trained model weights, evaluation scripts, publicly distributable external SRSs, and publicly distributable versions of ARID and SwaRD, has been made available via the ReqSeek GitHub repository (\url{https://github.com/kasrahabib/ReqSeek}). To support long-term access and citation, the same replication package has been archived on Zenodo and assigned a persistent identifier (\url{https://doi.org/10.5281/zenodo.24418048}). In addition to their availability through GitHub and Zenodo, ARID and SwaRD have also been made available on Hugging Face to facilitate easy download and use; they are accessible via repository IDs \texttt{kasrahabib/ARID} and \texttt{kasrahabib/SwaRD}, respectively.

Both ARID and SwaRD aggregate instances from source projects with heterogeneous licensing terms. Therefore, the released datasets are not distributed under a single unified license, and we do not re-license their constituent source datasets. Instead, source-specific attribution and license information are preserved in the dataset metadata, and users are responsible for complying with the applicable terms of the original sources. After loading a dataset, this information can be inspected through its metadata and accompanying documentation.

The experiments reported in this paper were conducted using the complete research datasets. The publicly distributed versions of ARID and SwaRD may exclude records from certain source projects due to restricted or unclear redistribution permissions. The current public-release composition and applicable restrictions are documented in the ``Public Release Notes'' section of the \texttt{datasets/README.md} file in the replication package. These redistribution restrictions do not affect the original experiments or the results reported in this study.

\bibliographystyle{ACM-Reference-Format}
\bibliography{software.bib}

\appendix

\section{Representative RQ4 Error-Characterization Examples}
\label{app:rq4_examples}

Table~\ref{tab:rq4_examples} presents representative examples from the qualitative characterization of RQ4 false-positive and false-negative cases. The examples originate from NTP and RMOS~\cite{senkyrSoftwareRequirementsData2023}, and KIB\textsuperscript{3}, which is released as part of this study's replication package. These examples are illustrative and are not used to recompute the primary RQ4 metrics, which remain based on the original SRS-derived labels. The complete false-positive/false-negative characterization file is included in the replication package: \url{https://github.com/kasrahabib/ReqSeek/blob/main/evaluations/RQ4/rq4_error_characterization.xlsx}.

\renewcommand{\arraystretch}{1.05}
\setlength{\tabcolsep}{4pt}
{\small
\begin{longtable}{@{}p{0.8cm} p{1.2cm} p{0.7cm} p{4.5cm} p{0.8cm} p{0.9cm} p{0.9cm} p{4cm}@{}}
\caption{Representative Examples from the Qualitative Characterization of RQ4 False-Positive and False-Negative Cases. Rows Are Color-Coded by Characterization Type. Abbreviations: OR = Omitted Requirement; FRP = False Requirement Prediction; MR = Missed Requirement; Pred. = ReqSeek Prediction; Insp. = Inspection Label Assigned by the First Author; AUX = Auxiliary; REQ = Requirement}
\label{tab:rq4_examples}\\
\toprule
\textbf{Char.} & \textbf{SRS} & \textbf{Err.} & \textbf{Sentence-level Unit} & \textbf{Gold} & \textbf{Pred.} & \textbf{Insp.} & \textbf{Insp. Note} \\
\midrule
\endfirsthead

\toprule
\textbf{Char.} & \textbf{SRS} & \textbf{Err.} & \textbf{Sentence-level unit} & \textbf{Gold} & \textbf{ReqSeek} & \textbf{Inspect.} & \textbf{Inspection note} \\
\midrule
\endhead

\midrule
\multicolumn{8}{r}{\textit{Continued on next page}}\\
\endfoot

\bottomrule
\endlastfoot

\rowcolor{omittederrcolor}
OR & KIB\textsuperscript{3} & FP & For the search functionality to work, a PDF file with (almost) the same name as the Moodle book must be present in the course, with each PDF page aligning one-to-one with the book pages. & AUX & REQ & REQ & Specifies a precondition/constraint for search functionality: a matching PDF file must be present and aligned. \\

\rowcolor{omittederrcolor}
OR & NTP & FP & The NPT should enable the Node Managers to customize the way GBIF-compliant data are served to their own community or to a specific group of data users. & AUX & REQ & REQ & States that NPT should enable node managers to customize how data are served; this is a system capability with modal force. \\

\rowcolor{omittederrcolor}
OR & NTP & FP & A user may only initiate the use cases in this section if he is logged into the system using a user account with administrator role. & AUX & REQ & REQ & Defines an access-control precondition for initiating use cases; this is a verifiable system constraint. \\

\rowcolor{omittederrcolor}
OR & NTP & FP & Post conditions The distribution map will use the projection and default extent the administrator chose. & AUX & REQ & REQ & Post-condition states how the distribution map will use selected projection and extent; this is expected system behavior. \\

\rowcolor{omittederrcolor}
OR & RMOS & FP & The payment system should be operable such that it can return information to the RMOS system as to whether payment was successful or failed. & AUX & REQ & REQ & Specifies that the payment system should return success/failure information to RMOS; this is an interface behavior. \\

\rowcolor{omittederrcolor}
OR & RMOS & FP & Users will summon their restaurant menu, which is combined with a system/command menu, using an easy touch gesture, a double-tap on the touch surface, and dismiss it with a similar gesture or by tapping a close button. & AUX & REQ & REQ & Describes the menu interaction gesture and dismissal behavior expected by users; this is a UI capability. \\

\rowcolor{omittederrcolor}
OR & RMOS & FP & The GUI will take a small percentage of the table's screen, so the UI will be clear and uncluttered. & AUX & REQ & REQ & Specifies GUI size and clarity constraints; this is a UI quality/property requirement. \\

\rowcolor{falseprederrcolor}
FRP & KIB\textsuperscript{3} & FP & The specialists and managers required for this transformation process must be trained early. & AUX & REQ & AUX & States that specialists and managers must be trained; this concerns people/training, not the software system. \\

\rowcolor{falseprederrcolor}
FRP & NTP & FP & - Green: NPT V1.0 must have this feature & AUX & REQ & AUX & Legend/priority marker explaining what green means; the concrete feature is not specified in this segment. \\

\rowcolor{falseprederrcolor}
FRP & RMOS & FP & The primary audience of this SRS document will be the development team employed to implement the specified RMOS. & AUX & REQ & AUX & Identifies the SRS audience; it does not specify system behavior or constraints. \\

\rowcolor{missederrcolor}
MR & KIB\textsuperscript{3} & FN & Basic The Python programming environment shall run in a safe manner, i.e., users shall not be able to execute malicious code. DONE & REQ & AUX & REQ & States the Python environment shall run safely and block malicious code; clear security/safety requirement. \\

\rowcolor{missederrcolor}
MR & KIB\textsuperscript{3} & FN & Basic Functional Integrating the English language beside German only for the software interface is a nice functionality to have. DONE & REQ & AUX & REQ & Describes a desired interface capability: integrating English beside German; requirement-like despite weak wording. \\

\end{longtable}
}

\section{Study Datasets}
We use the Hugging Face \texttt{datasets} API to distribute our datasets. This format supports efficient access to large-scale datasets, provides built-in operations such as \texttt{filter}, \texttt{map}, and streaming, and enables straightforward conversion to common formats such as \texttt{pandas}, \texttt{.csv}, and \texttt{.json}. These features facilitate reuse and make the datasets easier to integrate into downstream workflows. Hugging Face datasets can also be loaded across multiple programming environments, including Python, JavaScript, and Rust. In addition, the Hugging Face \texttt{datasets} API enables users to download and load the datasets directly into Python with minimal effort. Instructions for downloading, loading, inspecting, licensing, and using the datasets are provided in the dataset README of the replication package at \url{https://github.com/kasrahabib/ReqSeek/blob/main/datasets/README.md}

\subsection{Automatic Requirements Identification Dataset (ARID)}\label{apx:arid}
ARID is a manually curated and labeled dataset of ISO 29148-style requirements and auxiliary sentences collected from 23 publicly available SRSs and requirements datasets. ReqSeek was trained and evaluated on the full ARID dataset, which contains 2,396 instances: 1,050 requirement, 600 contextual\_auxiliary, and 746 system\_related\_auxiliary. Details on dataset availability and licensing are provided in Section~\ref{sec:da}.

\subsubsection{ARID Column Descriptions}
Table~\ref{tab:arid-columns} reports the metadata and column schema of the ARID dataset.

\begin{table}[h]
\centering
\caption{ARID Meta Information and Column Descriptions}
\label{tab:arid-columns}
\resizebox{\linewidth}{!}{%
\begin{tblr}{
  hline{1,10} = {-}{0.08em},
  hline{2} = {-}{0.05em},
}
\textbf{Column Name} & \textbf{Description}\\
\texttt{REQID} & Internal identifier assigned by the authors; may contain duplicate values.\\
\texttt{REQID\_expanded} & Unique identifier for each requirement in ARID.\\
\texttt{Project Name} & Name of the project from which the requirement originates.\\
\texttt{Subproject Name} & Name of a subproject within the larger project.\\
\texttt{Requirement Sentences} & Requirement text.\\
\texttt{signal\_keyword} & Manually labeled signal keyword in the sentence, e.g., \textit{shall}, \textit{should}, \textit{will}, \textit{may}, \textit{must}.\\
\texttt{binary\_cls} & Binary annotation: requirement vs. auxiliary, merging system-related and contextual auxiliary.\\
\texttt{ternary\_cls} & Multi-class annotation: requirement vs. system-related auxiliary vs. contextual auxiliary.
\end{tblr}
}
\end{table}

\subsubsection{ARID Dataset Examples}
To support transparency and enable qualitative inspection of the first author's manual labels, Table~\ref{tab:app-arid-samples} reports sample records from all three ARID classes. The examples below originate from ReqExp~\citep{ivanovReqExpBERTbasedML2021}, Crowd-Annotation Results~\citep{vanvlietCrowdAnnotationResultsIdentifying2020}, Dronology~\citep{cleland-huangRequirementsdrivenConfigurationEmergency2020}, NLP4RE ReqEval~\citep{Frieden84Nlp4rereqevalV20a}, ACTIVAGE~\citep{salaACTIVAGEUserNeeds_requirements2021}, and FN-RE~\citep{alhoshanFNRECorpusRequirements2018}.

\renewcommand{\arraystretch}{1.05}
\setlength{\tabcolsep}{4pt}
{\small
\begin{longtable}{@{}r p{3.6cm} p{8.2cm} p{2.8cm}@{}}
\label{app:arid-samples}\\
\caption{Sample Records from the ARID Dataset. \texttt{REQID\_expanded} is Abbreviated as \texttt{ID} for Readability. Rows are Color-Coded by Class Type}\label{tab:app-arid-samples}\\
\toprule
\textbf{ID} & \textbf{Project Name} & \textbf{Requirement Sentence} & \textbf{\shortstack{First Author\\Annotated Class}} \\
\midrule
\endfirsthead

\toprule
\textbf{ID} & \textbf{Project Name} & \textbf{Requirement Sentence} & \textbf{\shortstack{First Author\\Annotated Class}} \\
\midrule
\endhead

\midrule
\multicolumn{4}{r}{\textit{Continued on next page}}\\
\endfoot

\bottomrule
\endlastfoot
\rowcolor{ctxcolor}
35741 & \RaggedRight ReqExp & \RaggedRight Fastest lap in Italics. & contextual\_auxiliary \\
\rowcolor{ctxcolor}
27567 & \RaggedRight Crowd-Annotation Results: Identifying and Classifying User Requirements in Online Feedback & \RaggedRight They will not be able to make good decisions. & contextual\_auxiliary \\
\rowcolor{ctxcolor}
4024 & \RaggedRight Dronology & \RaggedRight In this case, consumers will replace one good with the other. & contextual\_auxiliary \\
\rowcolor{ctxcolor}
4236 & \RaggedRight Dronology & \RaggedRight Marshall Herff ``Do'' Applewhite Jr.May 17, 1931 – March 26, 1997) was an American cult leader. & contextual\_auxiliary \\
\rowcolor{ctxcolor}
4096 & \RaggedRight Dronology & \RaggedRight The European Union has made a few symbols that warn of dangers and that should be easy to understand. & contextual\_auxiliary \\
\rowcolor{reqcolor}
5342 & \RaggedRight NLP4RE ReqEval & \RaggedRight Where network services are chargeable, it should be possible for the network to provide information about call rates and on-going call charges. & requirement \\
\rowcolor{reqcolor}
359 & \RaggedRight ACTIVAGE & \RaggedRight Data should link to GP record so can view activity, plus also can gather info on hosital admissions etc. & requirement \\
\rowcolor{reqcolor}
4546 & \RaggedRight Dronology & \RaggedRight The \texttt{\_SingleUAVFlightPlanScheduler\_} shall maintain a managed list of pending flight plans for Each UAV. & requirement \\
\rowcolor{reqcolor}
4777 & \RaggedRight FN-RE & \RaggedRight The system will receive a notification when the door is closed. & requirement \\
\rowcolor{reqcolor}
4834 & \RaggedRight FN-RE & \RaggedRight The application should run on several operating systems, including at least Windows (XP and perhaps Vista), Linux and MacOS X. & requirement \\
\rowcolor{syscolor}
5215 & \RaggedRight NLP4RE ReqEval & \RaggedRight referential\textgreater{}It is possible that the initial NLM Repository may be composed of more than one electronic storage facility, especially in the early development of the repository. & system\_related\_auxiliary \\
\rowcolor{syscolor}
5238 & \RaggedRight NLP4RE ReqEval & \RaggedRight Each type of equipment has slightly different requirements placed upon it in terms of EMC and climate. & system\_related\_auxiliary \\
\rowcolor{syscolor}
4773 & \RaggedRight FN-RE & \RaggedRight Requests are definitive, i.e.they cannot be cancelled, and they persist; thus they should eventually be serviced. & system\_related\_auxiliary \\
\rowcolor{syscolor}
4615 & \RaggedRight Dronology & \RaggedRight The flight route that the UAV is going to cover is shown in regular lines. & system\_related\_auxiliary \\
\rowcolor{syscolor}
4559 & \RaggedRight Dronology & \RaggedRight An N-vector is a normal vector on the surface of the WGS-84 reference ellipsoid. & system\_related\_auxiliary \\

\end{longtable}
}

\subsection{Software Requirements Dataset (SwaRD)}\label{apx:sward}
\noindent SwaRD is constructed using our proposed ReqSeek model for automatic identification of requirements. SwaRD consists of 40,522 sentence-level instances from 231 publicly available SRSs and datasets. It comprises three classes: 10,201 \texttt{requirement}, 4,575 \texttt{contextual\_auxiliary}, and 25,746 \texttt{system\_related\_auxiliary} instances. Details on dataset availability and licensing are provided in Section~\ref{sec:da}.

\subsubsection{SwaRD Column Descriptions}
Table~\ref{tab:sward-columns} reports the metadata and column schema of the SwaRD dataset.

\begin{table}[h]
\centering
\caption{SwaRD Meta Information and Column Descriptions}
\label{tab:sward-columns}
\resizebox{\linewidth}{!}{%
\begin{tblr}{
  hline{1,12} = {-}{0.08em},
  hline{2} = {-}{0.05em},
}
\textbf{Column Name} & \textbf{Description}\\
\texttt{REQID} & Internal identifier assigned by the authors; may contain duplicate values.\\
\texttt{REQID\_expanded} & Unique identifier for each instance in SwaRD.\\
\texttt{Project Name} & Name of the project from which the sentence originates.\\
\texttt{Subproject Name} & Name of a subproject within the larger project.\\
\texttt{Requirement Sentences} & Requirement text.\\
$^*$\texttt{F/NF Classes} & Placeholder for functional (F) versus non-functional (NF) requirement classification.\\
$^*$\texttt{NF Subclasses} & Placeholder for non-functional requirement subclasses (e.g., performance, usability).\\
\texttt{isReqSysAuxContAux\_with\_keyword} & Signal keyword used (e.g., \textit{shall}, \textit{should}, \textit{will}, \textit{may}, \textit{must}).\\
\texttt{isReqSysAuxContAux} & Multi-class label: requirement vs. system-related auxiliary vs. contextual auxiliary.\\
\texttt{isReqAux} & Binary label: requirement vs. auxiliary (system/context merged).
\end{tblr}
}

\begin{minipage}{\linewidth}
\footnotesize
$^*$\textbf{Note:} The \texttt{F/NF Classes} and
\texttt{NF Subclasses} columns are unpopulated
placeholders (all values are \texttt{None})
intended for future requirements classification tasks.
\end{minipage}
\end{table}

\subsubsection{ReqSeek Identified Requirements in the SwaRD}
To support transparency and enable qualitative inspection of ReqSeek's identified labels, Table~\ref{tab:app-sward-samples} reports sample records from all three SwaRD classes. The examples below originate from A Dataset of Crowdsourced Smarthome Requirements with Creativity Ratings~\citep{murukannaiahAcquiringCreativeRequirements2016,MurukannaiahsmarthomerequirementsdatasetDataMaster}, IlmSeven~\citep{rathIlmSevenDataset2017a}, SecReq~\citep{knaussSecReq2021}, COEST~\citep{coest1}, PromiseExp~\citep{PROMISE_exparffa,limaSoftwareEngineeringRepositories2019a}, PURE~\citep{ferrariPUREDatasetPublic2018}, Requirements data sets (user stories)~\citep{dalpiazRequirementsDataSets2018}, ReqExp~\citep{ivanovReqExpBERTbasedML2021}, Crowd-Annotation Results~\citep{vanvlietCrowdAnnotationResultsIdentifying2020}, Dronology~\citep{cleland-huangRequirementsdrivenConfigurationEmergency2020}, NLP4RE ReqEval~\citep{Frieden84Nlp4rereqevalV20a}, ACTIVAGE~\citep{salaACTIVAGEUserNeeds_requirements2021}, and FN-RE~\citep{alhoshanFNRECorpusRequirements2018}.

\renewcommand{\arraystretch}{1.05}
\setlength{\tabcolsep}{4pt}
{\small
\begin{longtable}{@{}r p{3.6cm} p{8.2cm} p{2.8cm}@{}}
\label{app:sward-samples}\\
\caption{Sample Requirements Identified by ReqSeek in the SwaRD Dataset. \texttt{REQID\_expanded} is Abbreviated as \texttt{ID} for Readability. Rows are Color-Coded by Class Type}\label{tab:app-sward-samples}\\
\toprule
\textbf{ID} & \textbf{Project Name} & \textbf{Requirement Sentence} & \textbf{\shortstack{ReqSeek\\Identified Class}} \\
\midrule
\endfirsthead

\toprule
\textbf{ID} & \textbf{Project Name} & \textbf{Requirement Sentence} & \textbf{\shortstack{ReqSeek\\Identified Class}} \\
\midrule
\endhead

\midrule
\multicolumn{4}{r}{\textit{Continued on next page}}\\
\endfoot

\bottomrule
\endlastfoot
\rowcolor{ctxcolor}
38560 & \RaggedRight A Dataset of Crowdsourced Smarthome Requirements with Creativity Ratings & \RaggedRight It is the middle of the afternoon on a bright day. & contextual\_auxiliary \\
\rowcolor{ctxcolor}
36735 & \RaggedRight A Dataset of Crowdsourced Smarthome Requirements with Creativity Ratings & \RaggedRight However, it was just Daylights Saving Time so all the clocks changed by an hour. & contextual\_auxiliary \\
\rowcolor{ctxcolor}
24717 & \RaggedRight IlmSeven & \RaggedRight How do we tell which jobs failed and which didn't? & contextual\_auxiliary \\
\rowcolor{ctxcolor}
21655 & \RaggedRight IlmSeven & \RaggedRight The original idea of toolchains was to have the same tool being used by different plugins within the same project. & contextual\_auxiliary \\
\rowcolor{ctxcolor}
36837 & \RaggedRight A Dataset of Crowdsourced Smarthome Requirements with Creativity Ratings & \RaggedRight I get out of bed and walk toward the door. & contextual\_auxiliary \\
\rowcolor{reqcolor}
14365 & \RaggedRight IlmSeven & \RaggedRight The OpenFile dialog should show these files by default. & requirement \\
\rowcolor{reqcolor}
35847 & \RaggedRight SecReq & \RaggedRight When the CNG detects violation of the privacy protection policy the CNG shall discard all signalling, including signalling from the NGN. & requirement \\
\rowcolor{reqcolor}
3655  & \RaggedRight COEST & \RaggedRight Head-On/Passing Subservice shall be provided by an Advisory System. & requirement \\
\rowcolor{reqcolor}
35365 & \RaggedRight PromiseExp & \RaggedRight The system shall enable user to enter the payment information. & requirement \\
\rowcolor{reqcolor}
28892 & \RaggedRight PURE & \RaggedRight The System must be able to perform a simple search within 5--8 seconds and a advanced search (multiple search criteria) within 10--15 seconds regardless of the storage capacity or number of cases in the system. & requirement \\

\rowcolor{syscolor}
21721 & \RaggedRight IlmSeven & \RaggedRight Addition of a core extension point to the model builder supporting model finalization. & system\_related\_auxiliary \\
\rowcolor{syscolor}
9755  & \RaggedRight IlmSeven & \RaggedRight The tests AutoloadTest, JDBCDriversEmbeddedTest, JDBCDriversClientTest and JDBCDriversAllTest are in the junit-all ant target but not in suites. & system\_related\_auxiliary \\
\rowcolor{syscolor}
30749 & \RaggedRight PURE & \RaggedRight As such, the system should perform with consistently and predictably low response times in order not to impact the performance or the timely execution of the various tasks that need to be conducted by the Construction Junction staff. & system\_related\_auxiliary \\
\rowcolor{syscolor}
40508 & \RaggedRight Requirements data sets (user stories) & \RaggedRight As an archivist, I want to apply automatic contrast and level operations to a scanned file, so that I can eliminate unnecessary work. & system\_related\_auxiliary \\

\end{longtable}
}

\end{document}